\documentclass[aps,prl,twocolumn,superscriptaddress,floatfix,longbibliography,showpacs,nofootinbib]{revtex4-1}

\usepackage{graphicx}
\usepackage{dcolumn}
\usepackage{bm}
\usepackage{array}
\usepackage{amssymb}
\usepackage{microtype}
\usepackage{textcomp}
\usepackage{gensymb}
\usepackage{xcolor}
\usepackage{textalpha}
\usepackage{siunitx}
\usepackage{adjustbox}
\usepackage[justification = raggedright, singlelinecheck = false]{caption}
\usepackage{changes}
\usepackage[bottom]{footmisc}

\begin{document}

\title{Incommensurate magnetic orders and topological Hall effect in the square-net centrosymmetric EuGa$_2$Al$_2$ system}

\author{Jaime M. Moya}
\affiliation{Applied Physics Graduate Program, Smalley-Curl Institute, Rice University, Houston, Texas 77005, USA}
\affiliation{Department of Physics and Astronomy, Rice University, Houston, TX, 77005 USA}

\author{Shiming Lei}
\email{sl160@rice.edu}
\affiliation{Department of Physics and Astronomy, Rice University, Houston, TX, 77005 USA}

\author{Eleanor M. Clements}
\affiliation{NIST Center for Neutron Research, National Institute of Standards and Technology, Gaithersburg, MD 20899, USA}

\author{Caitlin S. Kengle}
\affiliation{Department of Physics, University of Illinois at Urbana–Champaign, Urbana, IL USA} 

\author{Stella Sun}
\affiliation{Department of Physics, University of Illinois at Urbana–Champaign, Urbana, IL USA} 

\author{Kevin~Allen}
\affiliation{Department of Physics and Astronomy, Rice University, Houston, TX, 77005 USA}

\author{Qizhi Li}
\affiliation{International Center for Quantum Materials, School of Physics, Peking University, CN-100871, Beijing, China}

\author{Y.Y. Peng}
\affiliation{International Center for Quantum Materials, School of Physics, Peking University, CN-100871, Beijing, China}

\author{Ali A. Husain}
\affiliation{Department of Physics and Astronomy and Quantum Matter Institute, University of British Columbia, Vancouver, British Columbia V6T 1Z1, Canada}

\author{Matteo Mitrano}
\affiliation{Department of Physics, Harvard University, Cambridge, Massachusetts 02138, USA}

\author{Matthew J. Krogstad}
\affiliation{Materials Science Division, Argonne National Laboratory, Lemont, IL, USA}

\author{Raymond Osborn}
\affiliation{Materials Science Division, Argonne National Laboratory, Lemont, IL, USA}

\author{Anand B. Puthirath}
\affiliation{Department of Materials Science and NanoEngineering, Rice University, Houston, TX, 77005 USA}

\author{Songxue Chi}
\affiliation{Neutron Scattering Division, Oak Ridge National Laboratory, Oak Ridge, Tennessee 37831, USA}

\author{L. Debeer-Schmitt}
\affiliation{Neutron Scattering Division, Oak Ridge National Laboratory, Oak Ridge, Tennessee 37831, USA}

\author{J. Gaudet}
\affiliation{NIST Center for Neutron Research, National Institute of Standards and Technology, Gaithersburg, MD 20899, USA}
\affiliation{Department of Materials Science and Eng., University of Maryland, College Park, MD 20742-2215}

\author{P. Abbamonte}
\affiliation{Department of Physics, University of Illinois at Urbana–Champaign, Urbana, IL USA}

\author{Jeffrey W. Lynn}
\affiliation{NIST Center for Neutron Research, National Institute of Standards and Technology, Gaithersburg, MD 20899, USA}

\author{E. Morosan}
\email{em11@rice.edu}
\affiliation{Department of Physics and Astronomy, Rice University, Houston, TX, 77005 USA}

\date{\today}

\begin{abstract}
%600 character limit 

Neutron diffraction on the centrosymmetric square-net magnet EuGa$_2$Al$_2$ reveals multiple incommensurate magnetic states (AFM1,2,3) in zero field. In applied field, a new magnetic phase (A) is identified from magnetization and transport measurements, bounded by two of the $\mu_0H$~=~0 incommensurate magnetic phases (AFM1,helical and AFM3, cycloidal) with different moment orientations. Moreover, magneto-transport measurements indicate the presence of a topological Hall effect, with maximum values centered in the A phase. Together, these results render EuGa$_2$Al$_2$ a material with non-coplanar or topological spin texture in applied field. X-ray diffraction reveals an out-of-plane (OOP) charge density wave (CDW) below $T_{CDW} \sim$ 50 K while the magnetic propagation vector lies in plane below $T_N$ = 19.5 K. Together these data point to a new route to realizing in-plane non-collinear spin textures through an OOP CDW. In turn, these non-collinear spin textures may be unstable against the formation of topological spin textures in an applied field.

\end{abstract}

\maketitle 

\renewcommand{\thefootnote}{\fnsymbol{footnote}}
\section{Introduction}
Magnetic skyrmions are particle-like spin textures of topological origin, which exist in the real space of materials \cite{nagaosa2013topological}. The intensive research on skyrmions has been driven by interest in their fundamental physical properties and potential applications to next-generation memory, logic, and neuromorphic computing devices \cite{fert2013skyrmions, fert2017magnetic, song2020skyrmion}. One prominent feature of magnetic skyrmions is that current densities required for the manipulation of their functionality are five to six orders of magnitude less than in modern spintronics \cite{jonietz2010spin,nagaosa2013topological}, establishing them as promising candidates for the design of energy-efficient electronic devices.

Following the experimental discovery of magnetic skyrmions in non-centrosymmetric crystals \cite{muhlbauer2009skyrmion, yu2011near, seki2012observation, adams2012long}, the existence of skyrmion lattices in centrosymmetric materials was quickly proposed theoretically and experimentally confirmed recently. Compared to non-centrosymmetric materials, where Dzyaloshinskii-Moriya (DM) interaction \cite{dzyaloshinsky1958thermodynamic, moriya1960new} is widely accepted as a critical driving force in stabilizing skyrmions, the mechanism for  skyrmion formation in centrosymmetric materials is less well understood. Theoretically, geometrically frustrated systems with short-range two-spin interactions were considered model candidates to host magnetic skyrmions \cite{okubo2012multiple,leonov2015multiply,lin2016ginzburg}. Alternatively, the four-spin interaction mediated by itinerant electrons has also been emphasized as an important ingredient \cite{heinze2011spontaneous,batista2016frustration, ozawa2017zero, hayami2017effective}. Lastly, skyrmionic bubbles in thin-film systems \cite{hou2017observation, xiao2020spin, he2022nanoscale} are a result of both intrinsic and extrinsic properties, but have distinct physics from bulk systems like EuGa$_2$Al$_2$ discussed here.

Despite substantive progress in theoretical understanding of skyrmion formation in centrosymmetric materials, the mechanism in real materials is still under debate. The limiting factor is the small number of known centrosymmetric skyrmion materials. So far, these have been limited to several intermetallic Gd-based compounds, including \mbox{Gd$_2$PdSi$_3$} \cite{kurumaji2019skyrmion,hirschberger2020topological} (triangular lattice), \mbox{Gd$_3$Ru$_4$Al$_{12}$} \cite{hirschberger2019skyrmion} (breathing kagom{\'e} lattice), and \mbox{GdRu$_2$Si$_2$} \cite{khanh2020nanometric, yasui2020imaging} (square lattice), and the perovskite oxide SrFeO$_3$ (cubic lattice) \cite{ishiwata2011versatile,ishiwata2020emergent}. Particularly, the experimental discovery of \mbox{GdRu$_2$Si$_2$} and SrFeO$_3$ as skyrmion hosts indicates that \textit{geometric} frustration may not be a necessary ingredient for the stabilization of skyrmions in centrosymmetric materials. Following the discovery of skyrmions in GdRu$_2$Si$_2$, Nomoto \textit{et. al.} \cite{Nomoto2020formation} studied the formation mechanism of the helical spin structure in \mbox{GdRu$_2$Si$_2$} and \mbox{Gd$_2$PdSi$_3$} by first principle calculations, and concluded that the \textit{interorbital} frustration inherent to Gd ions is the origin of the incommensurate spin modulation. By comparison, Hayami and Motome's work \cite{hayami2020square} based on an effective spin model suggested that the interplay of the four-spin interaction, bond-dependent anisotropic interaction and easy-axis anisotropy was essential for skyrmion formation in a square lattice. Independently, Wang \textit{et al}. \cite{wang2021meron} pointed out that multiple topological spin textures, including meron,
skyrmion, and vortex crystals, could be stabilized when only four-spin interactions and compass anisotropy were considered, without the need of a bare single-ion anisotropy. Clearly, it is paramount that new skyrmion-hosting materials, particularly non-Gd-based, are discovered in order to clarify the underlying mechanisms.

Here we report the observation of a nonzero topological Hall effect (THE) in the centrosymmetric square-lattice \mbox{EuGa$_2$Al$_2$}. The THE persists over a large range of magnetic field--temperature (\textit{H}--\textit{T}) space, peaking in an intermediate-field A phase. Neutron scattering measurements in zero field identify two magnetic phases (AFM1, AFM3) with non-collinear spin configurations and incommensurate propagation wavevectors ($\bf{q_{inc}}$). Another phase (AFM2) is identified with mixed wave vector separating AFM1 and AFM3. When a magnetic field is applied $H~\parallel~c$, the transition between AFM1 (with helical spin configuration and incommensurate propagation wavevector along the $a^*$ axis) and AFM3 (with cycloidal spin texture and incommensurate propagation vector along $a^*$) occurs \textit{via} the newly-discovered A phase, which may be a non-coplanar spin texture or a skyrmion state evidence by the observed THE. Furthermore, X-ray diffraction (XRD) measurements on \mbox{EuGa$_2$Al$_2$} reveal an out-of-plane (OOP) charge density wave (CDW) above T$_N$ (preformed CDW) which persists below T$_N$. Our combined magneto-transport and structural characterization in \mbox{EuGa$_2$Al$_2$} points to the important role of the preformed CDW in the formation of non-collinear spin textures in centrosymmetric materials. 

\section{Methods}

\begin{figure*}[t!]
  \includegraphics[width=\linewidth]{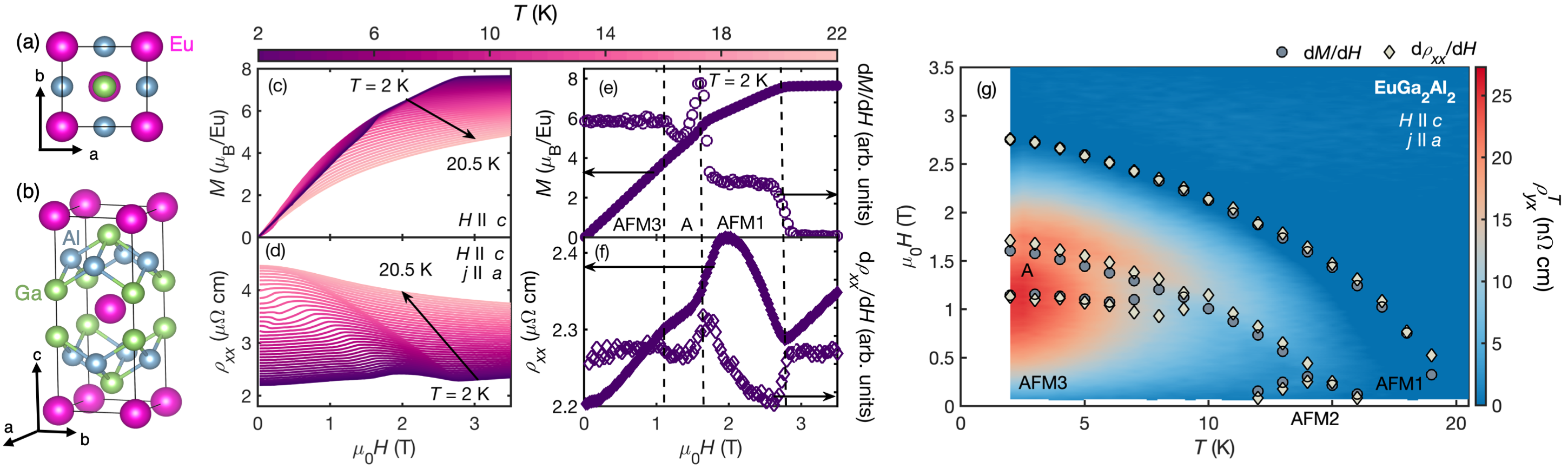}
\caption{Structure of \mbox{EuGa$_2$Al$_2$} and low temperature magnetic transitions. (a) and (b) top view and side view of \mbox{EuGa$_2$Al$_2$} structure.  Isothermal (c) magnetization $M$ and (d) resistivity $\rho_{xx}$ measured between $T = 2$~K (dark purple) and 20.5 K (light pink) with magnetic field  $H\parallel c$.  (e) $M$ measured at $T = 2$~K (left axis, closed circles) and the corresponding differential magnetic susceptibility d$M$/d$H$ (right axis, open circles). (f) $\rho_{xx}$ measured at $T = 2$~K (left axis, closed diamonds) and the corresponding derivative d$\rho_{xx}$/d$H$ (right axis, open diamonds). (g) The magnetic phase diagram constructed by local maxima in d$M$/d$H$ obtained from data in (c) (circles) and d$\rho_{xx}$/d$H$ from data in (e) (diamonds) together with the contour map of $\rho_{yx}^T$ established from Fig. \ref{transport}c with current $j \parallel a$.}.
\label{Aphase}
\end{figure*}

Single crystals of EuGa$_2$Al$_2$, EuGa$_4$ and EuAl$_4$ were grown by a self flux method described in Ref. \cite{stavinoha2018charge}. AC electrical transport measurements were performed in a Quantum Design (QD) DynaCool equipped with the Electrical Transport Option (ETO) and magnetization measurements were taken in a QD Dynacool with vibrating sample magnetometer option (VSM).  The tetragonal crystal symmetry dictates that the crystollographic $a$ and $b$ are equivalent. Throughout the manuscript, current $j \parallel a$ is applied while $H \parallel~c$. All contact resistances were measured to be less than 2 $\Omega$ at room temperature. Typical measurement parameters used were current $j~=~5~$mA and frequency $f~=~9.15~$Hz. The as-measured field-dependent resistivity was symmetrized and the corresponding Hall resistivity $\rho_{yx}$ was antisymmetrized. 

Magnetization as a function of magnetic field  was measured by sweeping the field from $0~\rightarrow~3.5~$T then back $3.5~\rightarrow~0~$T. No observable hysteresis was observed in magnetization or transport measurements in any region of the phase diagram.

The neutron scattering experiments were carried out on the HB-3 triple axis spectrometer at the High Flux Isotope Reactor at Oak Ridge National Laboratory.  The single crystal was in the shape of a flat plate of thickness 0.53 mm and weighed 48 mg.  A closed cycle refrigerator with a base temperature of 4.5 K was employed for the sample environment.  Most of the data were obtained using pyrolytic graphite (PG) filter and PG(002) monochromator at 35 meV incident and scattered neutron energies to reduce the very high absorption for Eu, and collimations of 48'-40'-40'-120' full-width-at-half-maximum (FWHM).  Data were collected in the ($h$,0,$l$), ($h$,$h$,$l$), and ($h$,$k$,0) scattering planes to search for other possible incommensurate magnetic peaks and to establish that the observed incommensurate wave vector is strictly along the [1,0,0] type direction.  A limited set of high resolution data were collected using the Si(111) monochromator with an energy of 9 meV and collimations of 48'-20'-20'-30' FWHM.  

In the ($h$,0,$l$) scattering plane, 27 magnetic Bragg reflections were collected as a function of rocking angle ($\theta$) at 5 K and 17 K to be used in the magnetic structure analysis of the high temperature AFM1 and low temperature AFM3 phases. Integrated intensities were corrected for instrument resolution and the heavy absorption of Eu. Absorption corrections were calculated assuming a flat plate geometry, where the tetragonal a-axis lies in the plane of the crystal face and the c-axis is normal to the plane. Correction ratios were rejected if the total path length for incoming and outgoing beams became longer than 0.5 times the crystal width ($w$ = 6 mm), viz. $r_i + r_f = r_T \geq 0.5 w$. Magnetic intensities were discarded from the refinement for interference with Al and Cu powder peaks, unphysical absorption corrections, and/or inability to measure a nearby nuclear Bragg peak for normalization. Refinements were carried out using the FullProf Suite \cite{rodriguez1993recent}. Wavevector $\bf{Q}$ and incommensurate reduced wavevector $q_{inc}$ are quoted in reciprocal lattice units (rlu) where $a^*$ = $2\pi/a$ and $c^*$ = $2\pi/c$ with $a$ = 4.3119 \AA and $c$ = 10.8939 \AA.

Small angle neutron scattering (SANS) measurements were performed at Oak Ridge National Laboratory on the CG-2 beamline using an 11 T horizontal field magnet as sample environment with a base temperature of 3 K. A single crystal sample of thickness 0.1 mm was mounted to achieve a scattering geometry with the magnetic field and sample c-axis nominally parallel to the incident neutron beam ($H~\parallel~c~\parallel \mathbf{k}_i$). Data were collected at a sample to detector distance of 2 m, an incident neutron wavelength $\lambda$ = 4 \AA, and resolution $\Delta\lambda/\lambda$ = 0.132. Measurements were collected via rocking scans satisfying the Bragg condition at magnetic fields up to 3 T.

X-ray diffraction measurements shown in the main text were performed with a low-emittance Xenocs GeniX 3D, Mo $K_{\alpha}$ (17.4 keV) source in an in-house set-up. A Huber four-circle diffractometer was used to control sample motion. A Mar345 image plate detector consisting of $12 \times 10^{6}$ pixels was used as a detector. The sample was cooled by closed-cycle cryostat with Be domes functioning as both vacuum and radiation shields to a base temperature of 8 K. 
	
A three-dimensional survey of momentum space was performed by moving the crystal through an angular range of $20^{\circ}$ in $\theta$ to index the crystal. After the CDW peak was located, a temperature-dependent series of three-dimensional surveys of momentum space were collected by moving the crystal through a smaller range of $\theta$ at sample temperatures between 8 K and 120 K.
	
The data collected were centered about the (1,1,2) Bragg peak. Line cuts were taken along the $L$ direction from 1.516 r.l.u.\ to 2.262 r.l.u.\ at each temperature. $H$ and $K$ values were accurate within $0.01$ r.l.u.
The intensities in Fig. \ref{CDW}b were integrated along the line cuts in Fig. \ref{CDW}a around each CDW peak after subtracting the background. 

All other X-ray diffraction data shown in the Supplementary materials were collected in the following way. Three-dimensional volumes of diffuse X-ray scattering were collected at the Advanced Photon Source on sector 6-ID-D using an incident energy of 87.1 keV on EuGa$_2$Al$_2$, EuGa$_4$, and EuAl$_4$. Sample temperatures from 300 K to 30~K were controlled by a nitrogen or helium gas flow. During the measurements, the samples were continuously rotated about an axis perpendicular to the beam at 1$\degree$ s$^{-1}$ over 360$\degree$, with images recorded every 0.1 s on a DECTRIS Pilatus 2M detector with a 1-mm-thick CdTe sensor layer. These were transformed to reciprocal space coordinates, allowing S(Q) to be determined over a range of $\pm$15 \AA$^{-1}$ in all directions. Further details are given in Ref. \cite{krogstad2020reciprocal}.

\section{Results and Discussion}
\subsection{Magnetic field - temperature phase diagram of EuGa$_2$Al$_2$}
\mbox{EuGa$_2$Al$_2$} is isostructural with the established skyrmion host GdRu$_2$Si$_2$ \cite{slaski1984magnetic, khanh2020nanometric, yasui2020imaging} (Fig. \ref{Aphase}a,b). Furthermore, the magnetism in the former originates in localised spin-only Eu$^{2+}$ ions (4f$^{7}$, $J = 7/2, L = 0$) \cite{stavinoha2018charge}, equivalent to Gd$^{3+}$ \cite{khanh2020nanometric}. The similarities between these two compounds extend to the complex magnetic order: EuGa$_2$Al$_2$ shows multiple magnetic transitions in zero field at $T_N~\approx~19.5$~K, $T_2~\approx~15$~K, and $T_3~\approx~11$~K \cite{stavinoha2018charge}. The lack of single-ion anisotropy and the multiple magnetic phases in zero field point to several competing energy scales, which are required for skyrmion formation. This observation in a non-Gd compound, and the structural and magnetic similarities with GdRu$_2$Si$_2$, motivated us to further determine the $H - T$ phase diagram in EuGa$_2$Al$_2$.

When an external magnetic field is applied along $c$ (the direction of the OOP CDW modulation, as discussed below), the magnetization measurements $M(H)$ (Fig.~\ref{Aphase}c) and resistivity measurements $\rho_{xx}(H)$ (Fig.~\ref{Aphase}d) reveal multiple transitions. Fig.~\ref{Aphase}e illustrates an $M(H)$ curve measured at $T~=~2~$K (full circles, left axis), where a new phase, not previously resolved (denoted as the A phase) is revealed in d$M$/d$H$ for $1.2$~T$~<~\mu_0H~<~1.6~$T (open circles, right axis). Evidence for the A phase is corroborated by $\rho_{xx}(H)$, Fig.~\ref{Aphase}f (full diamonds, left axis), and the derivative d$\rho_{xx}$/d$H$, Fig.~\ref{Aphase}f (open diamonds, right axis). The $H-T$ phase diagram of \mbox{EuGa$_2$Al$_2$} derived from magnetization $M(H)$ (circles) and resistivity $\rho_{xx}(H)$ (diamonds) is shown in Fig. \ref{Aphase}g. The three magnetic phases AFM1, AFM2, and AFM3 are consistent with previous reports \cite{stavinoha2018charge}, and the newly identified A phase is shown to exist in a narrow field range between $1.25$~T and $1.70$~T and temperatures up to $T~\approx~7$~K.

 \begin{figure}[t]
  \includegraphics[width=\linewidth]{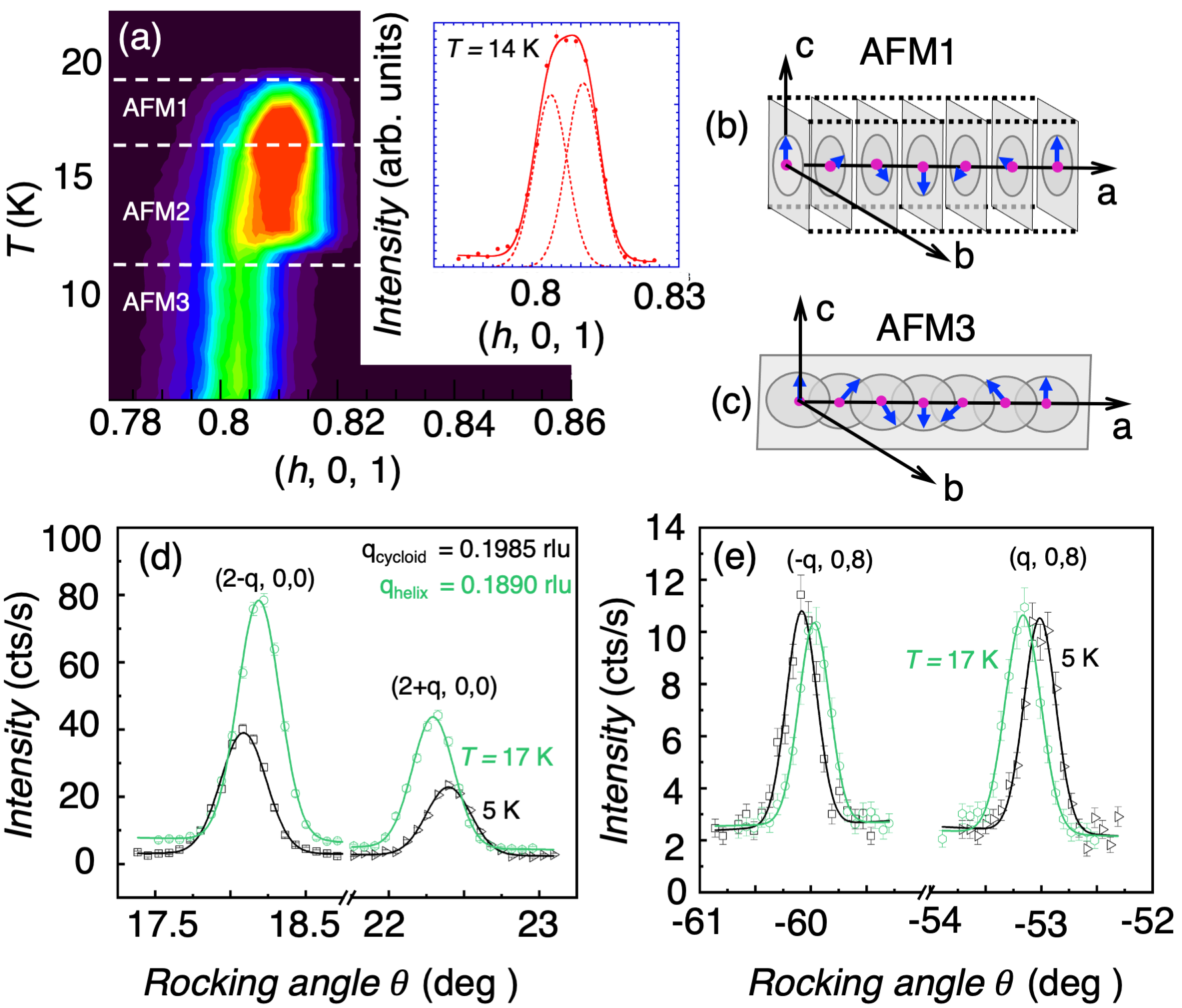}
\caption{Magnetic neutron diffraction for EuGa$_2$Al$_2$ in the ($h,0,l$) scattering plane. (a) Contour map of the diffraction intensity around the incommensurate magnetic Bragg reflection, $\bf{Q}$~=~($\sim$0.8,0,1) gives modulation wavevector $\pm \bf{q_{inc}}$ ($\sim$0.2,0,0). Inset: High resolution scan in the AFM2 regime shows substantial broadening due to scattering from two resolution-limited wave vectors in a mixed phase regime. Illustration of (b) helical and (c) cycloid magnetic phases of AFM1 and AFM3, respectively. (d) Satellites around the (2,0,0) fundamental Bragg peak approximately double in intensity from the AFM3 phase at 5 K to the AFM1 phase at 17 K, while (e) intensities of satellites at $\bf{Q}$ = ($\pm q_{inc}$,0,8) remain essentially unchanged between AFM1 and AFM3 phases.  Uncertainties, where indicated, represent one standard deviation. }
\label{neutron}
\end{figure}

\begin{figure}
  \includegraphics[width=\linewidth]{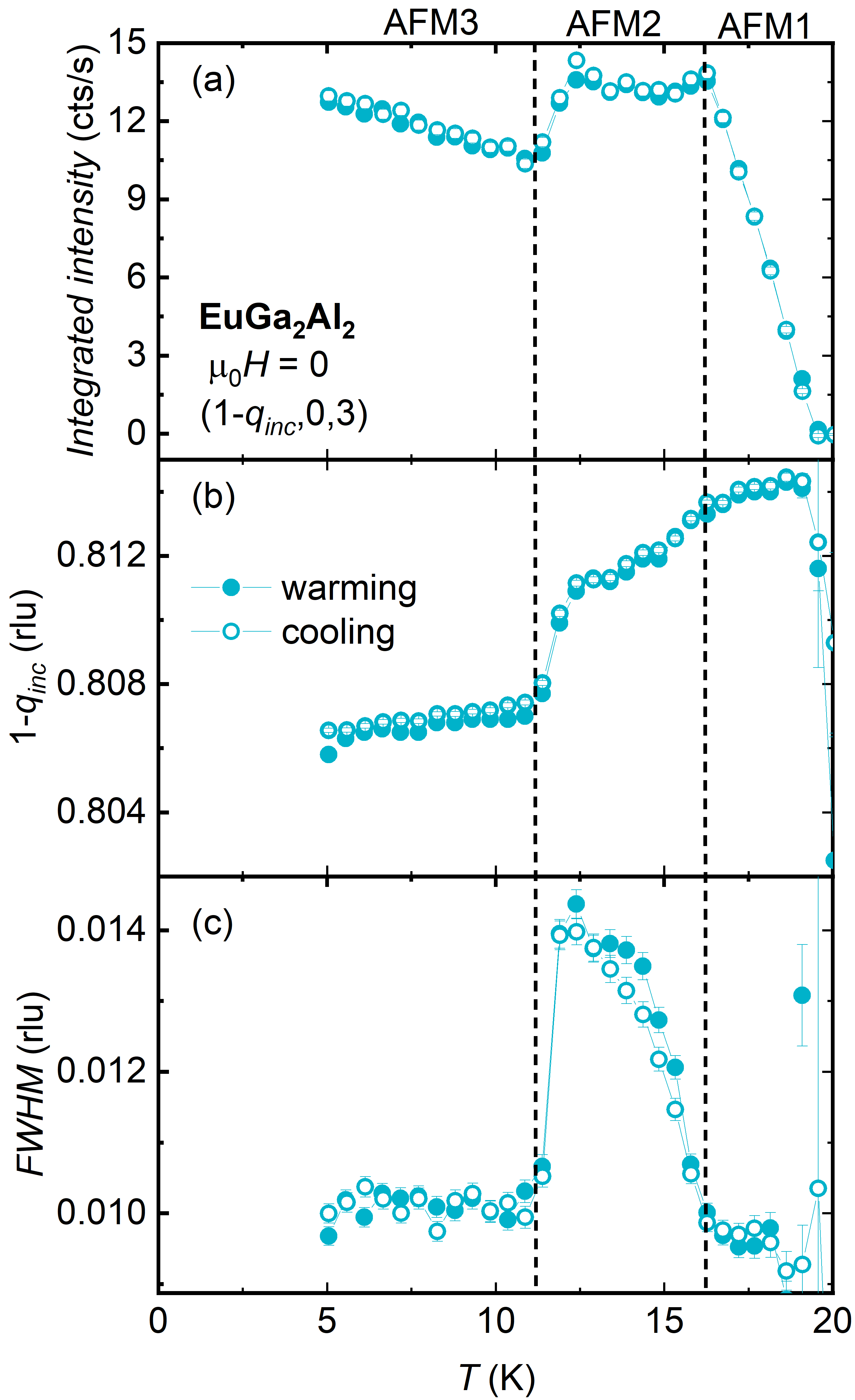}
\caption{Temperature dependence on warming and cooling of the incommensurate magnetic peak(s) near $\bm{Q}$ = (0.8,0,3) extracted using a single Gaussian fit to the data at each temperature. Lines are guides to the eye. (a) The integrated intensity, proportional to the square of the order parameter, identifies three distinct phases. A simple mean-field fit gives a transition temperature of $T_N$ = 19.5(2) K from the paramagnetic to AFM1 phase. (b) The wavevector peak position, 1-$q_{inc}$, and (c) full width at half maximum.  Uncertainties, where indicated, represent one standard deviation. }
\label{neutron2}
\end{figure}

Such a complex magnetic phase diagram is reminiscent of skyrmion host materials, where multiple non-collinear magnetic phases are typically found, a result of competing energy scales with comparable magnitudes \cite{tokura2020magnetic}. For example, in most non-centrosymmetric B20 skyrmion compounds, Heisenberg interactions, which favor parallel spin alignment, compete with DM interactions, which favor orthogonal spin alignment, resulting in non-collinear helical magnetic structures in zero field. Then the skyrmion spin structure can be realized with the application of a magnetic field. Therefore, a zero field non-collinear magnetic structure seems a necessary, albeit not sufficient, prerequisite for stabilizing a skyrmion state in applied field. The square-net compound GdRu$_2$Si$_2$ does show a zero field non-collinear incommensurate helical structure, and in this case a magnetic field has been shown to transform it into a skyrmion lattice \cite{khanh2020nanometric}.

\subsection{Elastic neutron scattering measurements}

To provide insight into the nature of the complex magnetism in EuGa$_2$Al$_2$, we turn to elastic neutron scattering experiments. The zero field neutron data in the ($h$,0,$l$) scattering plane shown in Fig. \ref{neutron}a indicate that, upon initially cooling into the ordered phase, an incommensurate helical state forms (AFM1) with $\bf{q_{inc}}$ along $a^*$ and the Eu moments rotating in the $bc$ plane as shown in Fig. \ref{neutron}b. At the lowest temperatures (AFM3), on the other hand, the ordering is again incommensurate, with a somewhat different value of the wavevector along $a^*$ (see  Fig. S1 in the Supplementary Materials at \cite{SM}), but forms a cycloid with the moments in the $ac$ plane as shown in  Fig. \ref{neutron}c. In the intermediate AFM2 phase, the contour map shows an increase and broadening in magnetic scattering intensity. Fig. \ref{neutron2} shows the details of the temperature dependence of the incommensurate wavevector.

A high resolution scan in the AFM2 regime can be fit with two resolution-limited peaks (inset Fig. \ref{neutron}a and Fig.~S1 located in the Supplementary Materials at \cite{SM}), indicating the presence of two wavevectors associated with a mixed state between the helical and cycloidal phases. The phase assignments at high and low temperatures are clearly indicated by the change in diffraction intensities between the AFM3 and AFM1 phases as shown in the scans in Fig. \ref{neutron}d,e. Recalling that neutrons only scatter from the component of the magnetization perpendicular to $\bf{Q}$, Fig. \ref{neutron}d compares the magnetic satellites at (2$\pm {q_{inc}}$,0,0) in the AFM1 and AFM3 phases. The intensities of both incommensurate peaks approximately double when the temperature is increased from 5 K to 17 K, indicating that there is magnetic scattering from an additional spin component perpendicular to $a^*$ in the AFM1 phase. In Fig. \ref{neutron}e, the intensities at ($\pm q_{inc}$,0,8) remain essentially unchanged between AFM1 and AFM3 when $\bf{Q}$ is essentially along the $c^*$ axis, which indicates no significant change in the single magnetic components scattering perpendicular to $c^*$. These characteristic changes in scattering intensities demonstrate that one component of the ordered moment rotates to the $a^*$-axis on cooling, consistent with a helical phase with moments in the $bc$ plane at 17 K transforming to a cycloid at 5 K with moments in the $ac$ plane, in both cases with the propagation vector along $a^*$. We note that another possibility is that a spin-density-wave forms initially below $T_N$.  In such a scenario, typically third and higher order peaks develop just below $T_N$  and grow in intensity \cite{choi2001direct}. We do not observe any such peaks in the AFM1 phase and thus have discarded this possibility. We find no evidence for an orthorhombic structural distortion below $T_N$, but any such distortion could be below our detection limit. Therefore we expect equivalent domains with propagation vectors along both $a^*$ and $b^*$.

\begin{figure}[t]
  \includegraphics[width=\linewidth]{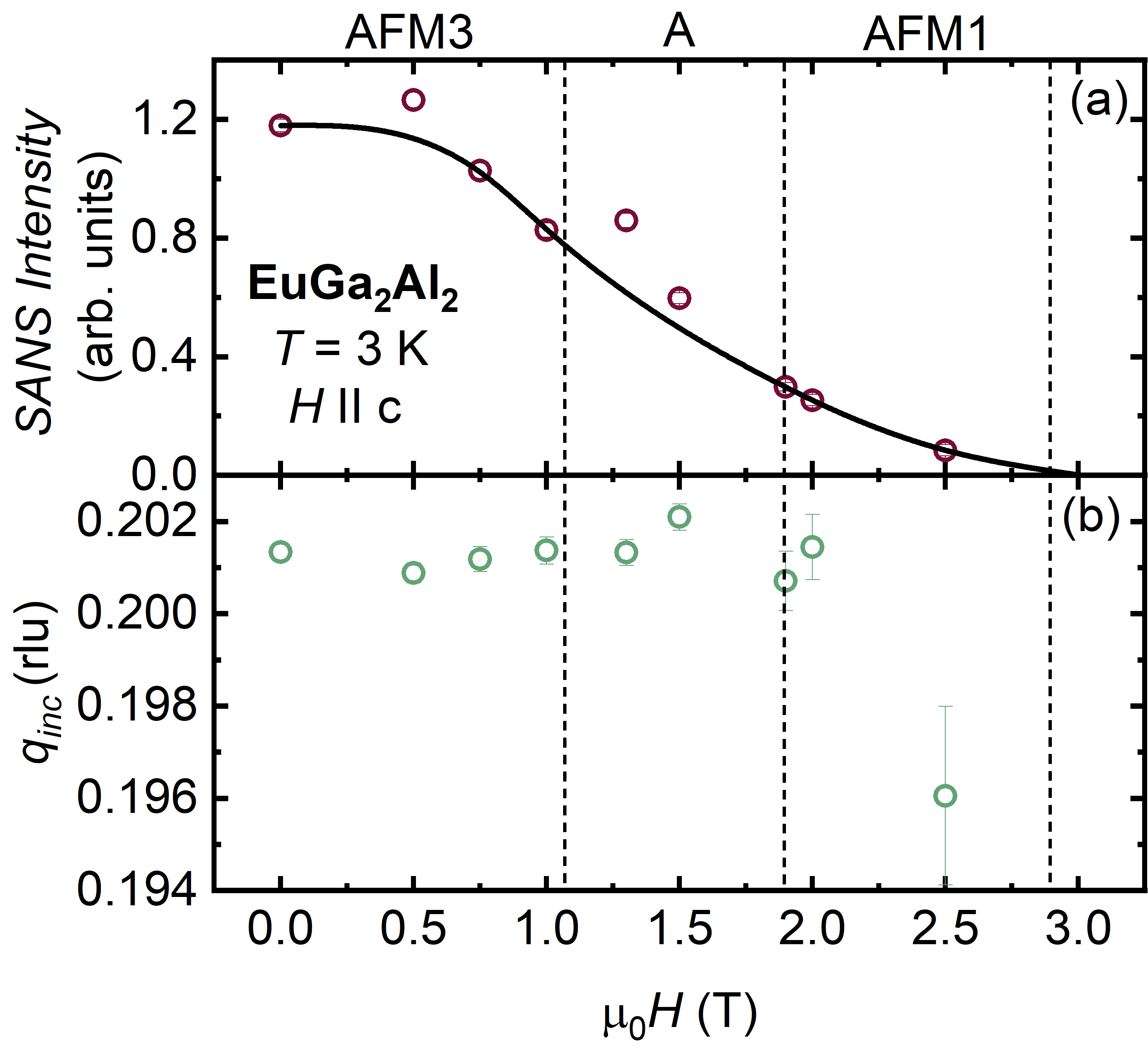}
\caption{Magnetic field dependence of the (a) integrated intensity and (b) position of the incommensurate wavevector, $q_{inc}$, extracted from Gaussian fits of the SANS intensity versus $Q_x$ shown in Fig.~S2 of the Supplementary Materials found at \cite{SM}.  The magnetic peak was unobservable at 3 T. Line is a guide to the eye. Error bars, where indicated, represent one standard deviation.}
\label{neutron3}
\end{figure}

We tracked the magnetic field dependence of the magnetic wavevector via SANS measurements. Fig.~ \ref{neutron3} shows the behavior of $q_{inc}$ extracted from Gaussian fits of the scattering intensity at -$Q_x$ on crossing from the AFM3 to A-phase to AFM1 for $H~\parallel~c$. The SANS intensity (Fig.~\ref{neutron3}a) generally falls off for fields applied along c and eventually disappears near 3 T, consistent with the transition to a field-polarized phase shown in the phase diagram (Fig.~\ref{Aphase}g) derived from magnetization and resistivity measurements.  In Fig.~\ref{neutron3}b, $q_{inc}$ appears to remain relatively constant across the three transitions. The wavevector appears to fall off above $H_{c2}$, however, the small intensity at 2.5 T introduces significant experimental uncertainty. Just above $H_{c1}~\sim$  1.1 T, we observe a kink in the SANS intensity (Fig.~\ref{neutron3}a) consistent with a phase transition across the AFM3 – A-phase boundary. The SANS diffraction pattern is shown in Fig.~S2 of the Supplementary Materials found at \cite{SM}, shows the four-fold pattern of the tetragonally equivalent magnetic peaks. However, our SANS survey of the ($h$,$k$,0) scattering plane did not show additional spots at $Q_1 + Q_2$, which were observed recently in EuAl$_4$ and, together with a polarization analysis interpreted as evidence for a skyrmion state \cite{takagi2022square}.

 If the THE in EuGa$_2$Al$_2$, discussed next, originates from skyrmions then one possibility is that they haven't formed a well ordered lattice in this crystal for some reason, or a well formed lattice does develop but with a different orientation, perhaps from the strong coupling to the c-axis CDW discussed below.  We cannot exclude the possibility of skyrmion lattice in EuGa$_2$Al$_2$, and further surveys of  reciprocal space together with a polarization analysis are needed to understand the exact nature of the A phase in this square lattice. Of course the THE in this material that we now discuss certainly may originate from a different topological spin texture.

\subsection{Topological Hall effect in EuGa$_2$Al$_2$}

\begin{figure}
  \includegraphics[width=\linewidth]{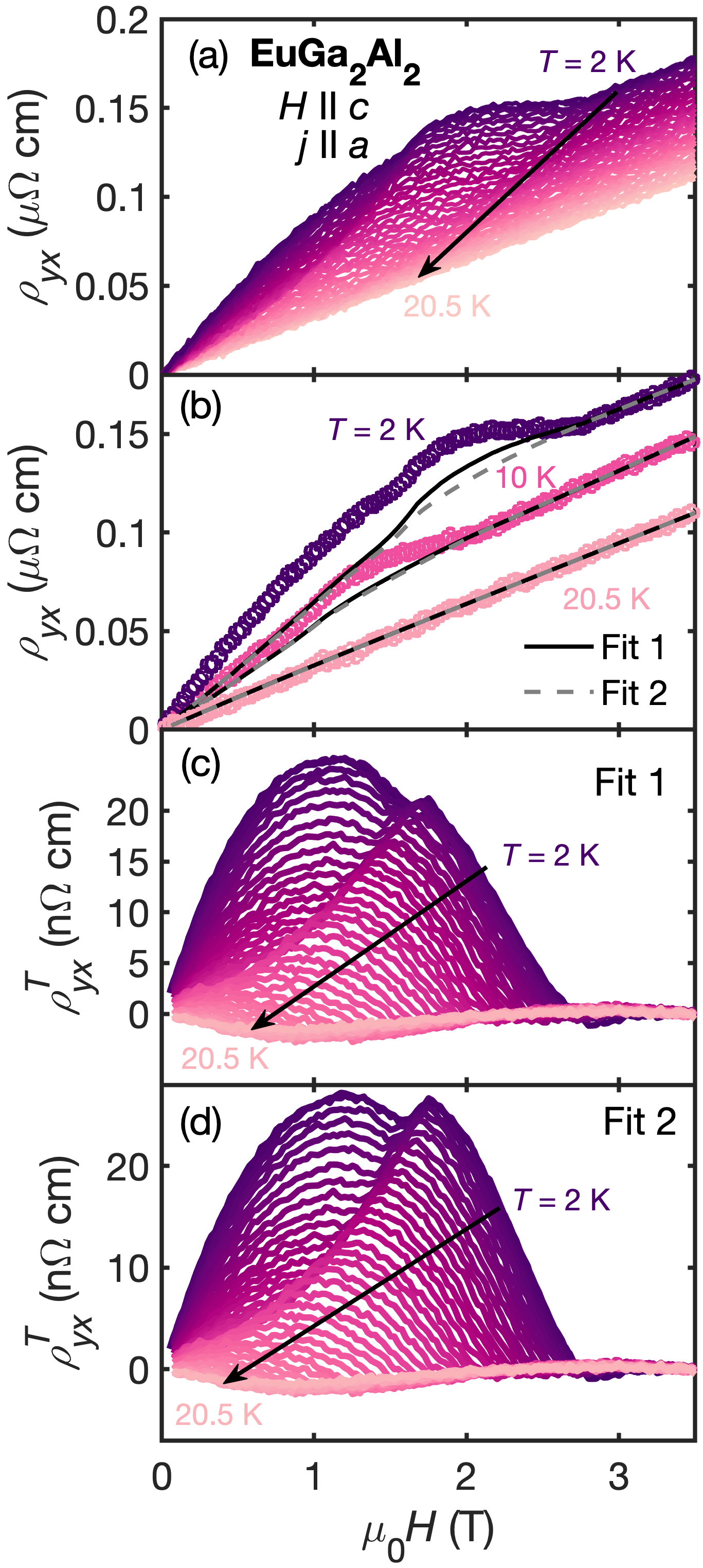}
\caption{Topological Hall effect in  \mbox{EuGa$_2$Al$_2$}. (a) Hall resistivity measurements measured between $T$ = 2 K (dark purple) and 20.5 K (light pink) with $H \parallel c$ and current $j \parallel a$. (b) A subset of the $\rho_{yx}$ data (symbols)  at $T = 2$ K (purple), 10 K (magenta), and 20.5 K (pink). Solid lines are fits to Eq. \ref{Fit 1} while dashed lines are fits to Eq.~\ref{Fit 2}.  The topological Hall resistivity $\rho_{yx}^T$ determined by subtracting the anomalous Hall resistivity $\rho_{yx}^A$ and normal Hall resistivity $R_0\mu_0H$ from $\rho_{yx}$ using the resultant fits from (c)  Eq. \ref{Fit 1} or (d)  Eq. \ref{Fit 2} from $T = 2$ K (dark purple) to 20.5 K (light pink) in 0.5 K increments. }
\label{transport}
\end{figure}

With AFM1 and AFM3 phases determined to be incommensurate non-collinear magnetic structures but with different moment orientation, the ncommensurate magnetic structure in the  A phase is likely a non-coplanar magnetic phase, with skyrmion lattice being a possibility. Non-coplanar spin textures have non-zero scalar spin chirality defined as $\chi=\bm{S_i}\cdot(\bm{S_j}\times\bm{S_k})$ where $\bm{S_i},\bm{S_j},\bm{S_k}$ are nearest neighbor spins \cite{ueda2012topological}. When itinerant electrons are coupled to such a non-coplanar spin texture, they acquire a Berry phase which is proportional to $\chi$. Therefore, the non-coplanar spin textures produce an effective magnetic field, and additional contributions to the Hall resistivity are expected \cite{tokura2020magnetic}.

We therefore turn to field-dependent electrical transport measurements to shed light on the nature of the A phase in \mbox{EuGa$_2$Al$_2$}. In Fig. \ref{transport}a we plot the temperature dependence of the measured Hall resistivity $\rho_{yx}(H)$ ($j\parallel a$, $H\parallel c$) while the $T$~=~2~K,~10~K,~and~20.5~K data are highlighted in Fig.~\ref{transport}b.  Non-linear $\rho_{yx}(H)$ is recorded in the ordered state ($T$ = 2~K and 10~K), with the Hall resistivity becoming virtually linear at $T$ = 20.5 K (in the paramagnetic state).

In the presence of a non-coplanar spin texture, the Hall resistivity can be expressed as the sum of several contributions: $\rho_{yx} = R_0\mu_0H + \rho_{yx}^A + \rho_{yx}^T$, where $R_0$ is the normal Hall coefficient, $\rho_{yx}^A$ the anomalous Hall resistivity, and $\rho_{yx}^T$ the THE resistivity. The anomalous contribution to the Hall resistivity, $\rho_{yx}^A$, can either be expressed as $S_H\rho_{xx}^2M$ for a dominant intrinsic scattering mechanism or $S_H'\rho_{xx}M$ for skew scattering. The intrinsic mechanism is expected to dominate in moderately disordered systems, while the skew scattering mechanism is expected to dominate in ultraclean systems \cite{nagaosa2010anomalous}. We compared the fits to the measured $\rho_{yx}$ data using both empirical expressions. For Fit 1  (Fig. \ref{transport}b, solid lines) we assume a dominant intrinsic mechanism, and fit the data in the spin polarized state where $M$ saturates, therefore $\rho_{yx}^T$ becomes negligible, and $\rho_{yx}$ is linear to

\begin{equation}
\frac{\rho_{yx}}{\mu_0H} = R_0 + \frac{S_H\rho_{xx}^2M}{\mu_0H}.
\label{Fit 1}
\end{equation}

Here, $R_0$ is extracted as the intercept and $S_H$ is the slope of the linear fit when plotting $\rho_{yx}/\mu_0H$ \textit{vs.} $S_H\rho_{xx}^2M/\mu_0H$. Similarly, for Fit 2 (Fig. \ref{transport}b, dashed lines), assuming skew scattering is the dominant mechanism, the same data in the field polarized regime are fit to 

\begin{equation}
\frac{\rho_{yx}}{\mu_0H} = R_0 + \frac{S_H'\rho_{xx}M}{\mu_0H}.
\label{Fit 2}
\end{equation}

The THE $\rho_{yx}^T$ is captured by the difference between the experimental data (symbols) and the fits (lines). The two fits are qualitatively very similar, with Fit 1, Fig.~\ref{transport}c, doing  slightly better at minimizing $\rho_{yx}^T$ compared to Fit 2 shown in (Fig.~\ref{transport}d).  It appears that THE is maximum around 1.2 T, in the A phase, as indicated by the contour plot in the $H - T$ phase diagram of Fit 1 (Fig. \ref{Aphase}g). This further supports the scenario of a non-coplanar spin texture in the A phase.

\subsection{Incommensurate, out-of-plane charge-density wave in EuGa$_2$Al$_2$ }
We now look into the  origin of non-collinear spin textures in centrosymmetric magnetic compounds in general, and in the square-net compound EuGa$_2$Al$_2$ in particular. In intermetallic compounds like EuGa$_2$Al$_2$, the Ruderman-Kittel-Kasuya-Yosida (RKKY) interaction \cite{ruderman1954indirect,kasuya1956theory, yosida1957magnetic} underlines the interplay of conduction electrons and local moments. This spin-charge interaction has been shown theoretically to induce a CDW from a spin-density state \cite{hayami2021charge}, and was confirmed experimentally with the observation of the in-plane charge-modulation in the skyrmion state of GdRu$_2$Si$_2$ \cite{yasui2020imaging}. However, this behavior is different in the case of EuGa$_2$Al$_2$, where the \textit{OOP} CDW is perpendicular to the in-plane magnetic wavevector. %\textit{preformed}  ($T_{CDW}>T_N$). 

\begin{figure}[t!]
 \includegraphics[width=\linewidth]{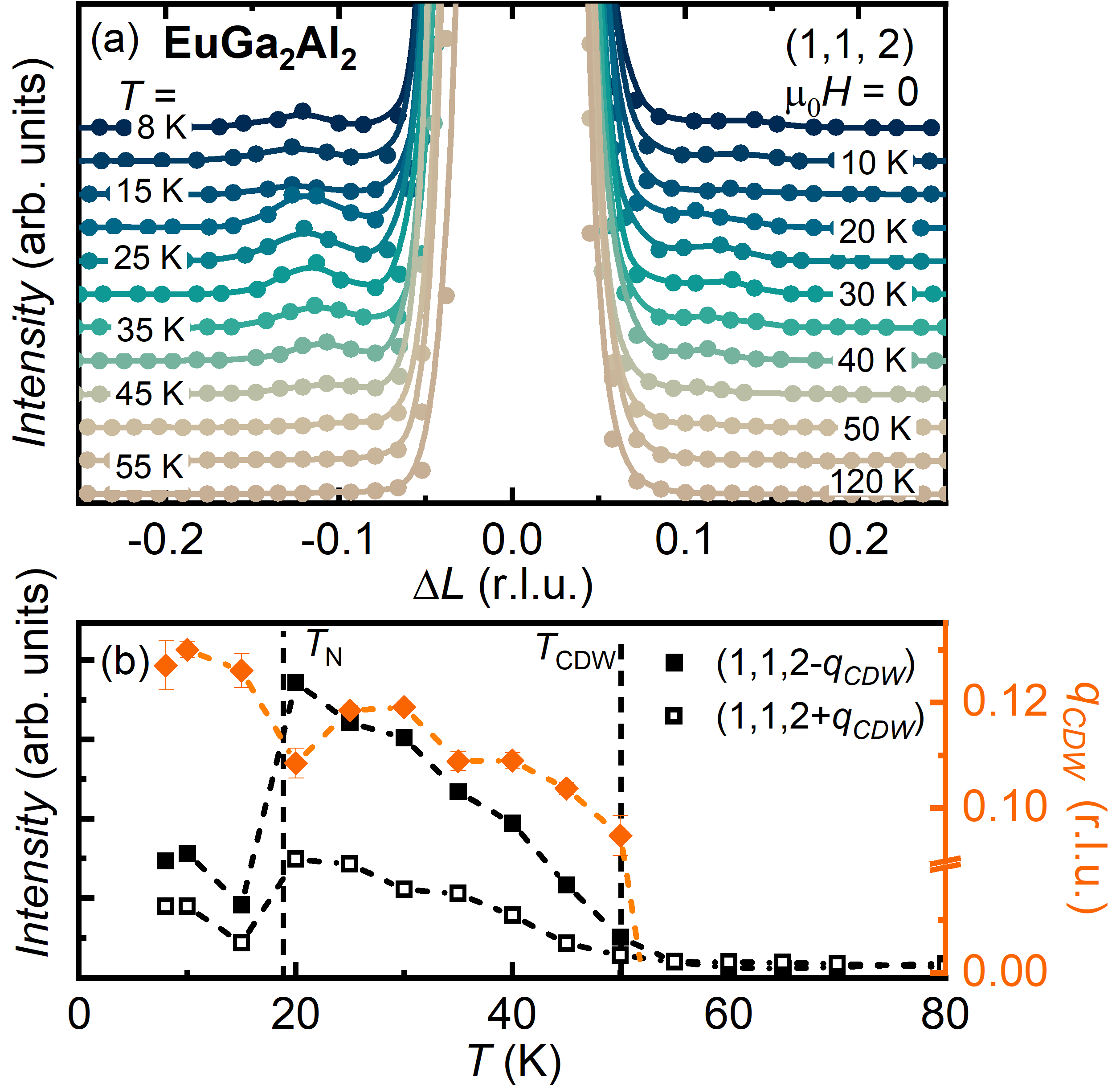}
\caption{XRD measurements in EuGa$_2$Al$_2$.  (a) Temperature dependent line-profile cuts along (1,~1,~2~$\pm~\Delta L$) for $8 \leq T \leq 120$ K, and (b) the temperature dependence of the order parameter defined as the integrated intensity of the peak with the background subtracted, located at (1,~1,~2~+(-)~$q_{CDW}$) (left axis, open (closed) squares) and the temperature dependence of $q_{CDW}$ (right axis, diamonds). The vertical dashed lines at $T_N$ and $T_{CDW}$ are determined from thermodynamic and transport measurements. Uncertainties, where indicated, represent one standard deviation.}
\label{CDW}
\end{figure}

Figure \ref{CDW} shows the temperature-dependent XRD measurements, which confirm that the $\sim$ 50 K anomaly previously observed in the resistivity in EuGa$_2$Al$_2$ \cite{stavinoha2018charge} is due to a CDW transition. The line-profile cuts of the XRD data around the (1,~1,~2) Bragg peak (Fig. \ref{CDW}a) reveal no superlattice reflections for $T~\geq~50~$K. At $T~<$ 50 K  clear superlattice reflections are observed, in agreement with the CDW transition temperature $T_{CDW}\approx~50$ K established from $\rho(T)$ \cite{stavinoha2018charge}. The expected second-order CDW peaks are clearly observed in the reciprocal space maps (Fig. S3 and S4 in the Supplementary Materials at \cite{SM}). The temperature dependence of the superlattice reflections located at (1,~1,~2~+(-)~$q_{CDW}$) (Fig. \ref{CDW}b, left axis, open (closed) squares) demonstrate that the amplitude of the reflections continuously grows before sharply decreasing below $T_N$, increasing on further cooling, indicating strong spin-charge coupling. The spin-charge coupling is corroborated by the temperature dependence of $q_{CDW}$ (Fig. \ref{CDW}b, right axis, diamonds), which increases from 0.09 reciprocal lattice units (r.l.u.) at 45 K to 0.12 r.l.u. at 25 K, before a sharp decrease is observed near $T_N$. $q_{CDW}$ again increases on further cooling. Our data thus demonstrate the presence of an incommensurate, preformed ($T_{CDW} > T_N$), OOP CDW state in EuGa$_2$Al$_2$, which persists through the magnetically ordered state with in-plane modulation vectors.

To understand the implications of the OOP CDW state on the magnetic order, we consider EuGa$_2$Al$_2$ in the context of the whole Eu(Ga$_{1-x}$Al$_x$)$_{4}$ series, and particularly by comparison with the two end members EuAl$_4$ and EuGa$_4$. The three structurally-ordered compounds EuAl$_4$, EuGa$_2$Al$_2$ and EuGa$_4$ feature Eu square-net planes separated by (Ga,Al)$_4$ layers. Previously, an OOP CDW transition in EuAl$_4$ had been reported by XRD and neutron measurements near $T_{CDW}$ $\approx~140~$K \cite{shimomura2019lattice,kaneko2021charge}. Furthermore, complex magnetism with successive magnetic transitions characterized by incommensurate ordering wave vectors had also been reported \cite{kaneko2021charge}. The temperature dependence of $q_{CDW}$ shows sharp inflections at T$_N$, indicative of strong spin-charge coupling in EuAl$_4$. Importantly, the OOP CDW persists in the $H = 0$ magnetically ordered phases \cite{shimomura2019lattice}. Although the exact magnetic structure for EuAl$_4$ has yet to be determined, non-zero THE has recently been reported for $H~\parallel~c$ \cite{shang2021anomalous}. By contrast, EuGa$_4$ is distinct from both EuGa$_2$Al$_2$ and EuAl$_4$, as it has only one magnetic transition into a simple collinear AFM structure characterized by a single commensurate propagation vector \cite{kawasaki2016magnetic}. No CDW or THE is observed in EuGa$_4$ at ambient pressure \cite{stavinoha2018charge, nakamura2015transport, Kevin2021}, as shown by the temperature-dependent XRD (see Fig. S5 in the Supplementary Material at \cite{SM}) and transport data \cite{Kevin2021}. Together, these results suggest that the preformed incommensurate OOP CDW, which persists through the magnetically ordered state in EuGa$_2$Al$_2$ and EuAl$_4$, is necessary for the formation of low temperature non-collinear magnetic textures. This is reinforced by the fact that, in the absence of a CDW in EuGa$_4$, the resulting magnetic state is a simple commensurate magnetic ordered state. 

The mechanism of OOP CDWs contributing to the formation of complex in-plane magnetism can be rationalized as follows. The incommensurate magnetic phases such as those observed in EuGa$_2$Al$_2$ are low-energy ordered states, and their instability can be related to the nesting properties of the Fermi surface in metals. The peak location of the bare magnetic susceptibility dictates the wave vector and thus the noncollinear spin texture.  In EuGa$_2$Al$_2$ (and EuAl$_4$), the preformed CDW above $T_N$ necessarily leads to a Fermi surface reconstruction and consequently modifies the landscape of the bare susceptibility. The OOP CDW results in band folding in the OOP direction which effectively reduces the dimensionality of the Fermi surface (\textit{i.e.} renders it more two-dimensional), leaving the system more susceptible to the formation of a noncollinear spin texture with in-plane propagation vector. This is in contrast to EuGa$_4$ with no CDW where the Fermi surface is necessarily more three-dimensional. Our results thus indicate that the OOP CDW in the square-net compounds favor the formation of non-collinear spin textures, which in turn set the stage for  non-coplanar spin textures with the application of a magnetic field.  

\section{Conclusions}
In conclusion, we determine the EuGa$_2$Al$_2$ magnetic phase diagram for $H \parallel c$, with three zero field magnetic phases present in the antiferromagnetic ordered state. With the application of a magnetic field $H \parallel c$, a new magnetic phase A emerges, separating AFM3 and AFM1.  Neutron diffraction measurements in zero field reveal a cycloid spin structure in AFM3 with an incommensurate propagation vector in the $a^*$ direction, while an incommensurate helical structure propagating in the same direction is observed in AFM1. The transition between these two non-collinear spin states takes place \textit{via} the A phase in applied  field, and gives rise to the observed THE which is maximized in this intermediate phase. Our XRD measurements reveal a preformed incommensurate OOP CDW propagating along the $c^*$ direction emerging below 50 K, which is strongly coupled to and persists into the magnetically ordered state. Taken in context with the isostructural compounds EuAl$_4$ and EuGa$_4$, our results point to the OOP CDW as a possible driving mechanism  in stabilizing non-collinear spin textures. Such non-collinear spin textures can,  in turn,  give way to non-coplanar or even skyrmion spin textures with the application of magnetic field.

\section{\label{sec:level1} ACKNOWLEDGMENTS}
JMM was supported by the National Science Foundation (NSF) Graduate Research Fellowship under Grant DGE 1842494. EM and SL acknowledge support from U.S. DOE BES DE- SC0019503. KJA was supported by NSF DMR 1903741. XRD measurements were supported by the U.S. Department of Energy, Office of Basic Energy Sciences Grant No. DE-FG02-06ER46285. P. A. acknowledges the Gordon and Betty Moore Foundation's EPiQS Initiative through Grants No. GBMF9452. This research used resources at the High Flux Isotope Reactor, Oak Ridge National Laboratory, and the Advanced Photon Source, Argonne National Laboratory, which are DOE Office of Science User Facilities supported by the US Department of Energy, Office of Science, Office of Basic Energy Sciences. Work at the Materials Science Division, Argonne National Laboratory (single crystal diffuse x-ray scattering) was supported by US Department of Energy, Office of Science, Office of Basic Energy Sciences, Materials Sciences and Engineering Division. The use of the EPMA facility at the Department of Earth Science, Rice University, Houston, Texas, is kindly acknowledged. The identification of any commercial product or trade name does not imply endorsement or recommendation by the National Institute of Standards and Technology.\\

During the review of this manuscript, a skyrmion state was reported in EuAl$_4$ \cite{takagi2022square}.

\bibliography{EuGa2Al2_finalarxiv}

%merlin.mbs apsrev4-1.bst 2010-07-25 4.21a (PWD, AO, DPC) hacked
%Control: key (0)
%Control: author (72) initials jnrlst
%Control: editor formatted (1) identically to author
%Control: production of article title (-1) disabled
%Control: page (0) single
%Control: year (1) truncated
%Control: production of eprint (0) enabled
\begin{thebibliography}{51}%
\makeatletter
\providecommand \@ifxundefined [1]{%
 \@ifx{#1\undefined}
}%
\providecommand \@ifnum [1]{%
 \ifnum #1\expandafter \@firstoftwo
 \else \expandafter \@secondoftwo
 \fi
}%
\providecommand \@ifx [1]{%
 \ifx #1\expandafter \@firstoftwo
 \else \expandafter \@secondoftwo
 \fi
}%
\providecommand \natexlab [1]{#1}%
\providecommand \enquote  [1]{``#1''}%
\providecommand \bibnamefont  [1]{#1}%
\providecommand \bibfnamefont [1]{#1}%
\providecommand \citenamefont [1]{#1}%
\providecommand \href@noop [0]{\@secondoftwo}%
\providecommand \href [0]{\begingroup \@sanitize@url \@href}%
\providecommand \@href[1]{\@@startlink{#1}\@@href}%
\providecommand \@@href[1]{\endgroup#1\@@endlink}%
\providecommand \@sanitize@url [0]{\catcode `\\12\catcode `\$12\catcode
  `\&12\catcode `\#12\catcode `\^12\catcode `\_12\catcode `\%12\relax}%
\providecommand \@@startlink[1]{}%
\providecommand \@@endlink[0]{}%
\providecommand \url  [0]{\begingroup\@sanitize@url \@url }%
\providecommand \@url [1]{\endgroup\@href {#1}{\urlprefix }}%
\providecommand \urlprefix  [0]{URL }%
\providecommand \Eprint [0]{\href }%
\providecommand \doibase [0]{http://dx.doi.org/}%
\providecommand \selectlanguage [0]{\@gobble}%
\providecommand \bibinfo  [0]{\@secondoftwo}%
\providecommand \bibfield  [0]{\@secondoftwo}%
\providecommand \translation [1]{[#1]}%
\providecommand \BibitemOpen [0]{}%
\providecommand \bibitemStop [0]{}%
\providecommand \bibitemNoStop [0]{.\EOS\space}%
\providecommand \EOS [0]{\spacefactor3000\relax}%
\providecommand \BibitemShut  [1]{\csname bibitem#1\endcsname}%
\let\auto@bib@innerbib\@empty
%</preamble>
\bibitem [{\citenamefont {Nagaosa}\ and\ \citenamefont
  {Tokura}(2013)}]{nagaosa2013topological}%
  \BibitemOpen
  \bibfield  {author} {\bibinfo {author} {\bibfnamefont {N.}~\bibnamefont
  {Nagaosa}}\ and\ \bibinfo {author} {\bibfnamefont {Y.}~\bibnamefont
  {Tokura}},\ }\href@noop {} {\bibfield  {journal} {\bibinfo  {journal} {Nature
  nanotechnology}\ }\textbf {\bibinfo {volume} {8}},\ \bibinfo {pages} {899}
  (\bibinfo {year} {2013})}\BibitemShut {NoStop}%
\bibitem [{\citenamefont {Fert}\ \emph {et~al.}(2013)\citenamefont {Fert},
  \citenamefont {Cros},\ and\ \citenamefont {Sampaio}}]{fert2013skyrmions}%
  \BibitemOpen
  \bibfield  {author} {\bibinfo {author} {\bibfnamefont {A.}~\bibnamefont
  {Fert}}, \bibinfo {author} {\bibfnamefont {V.}~\bibnamefont {Cros}}, \ and\
  \bibinfo {author} {\bibfnamefont {J.}~\bibnamefont {Sampaio}},\ }\href@noop
  {} {\bibfield  {journal} {\bibinfo  {journal} {Nature nanotechnology}\
  }\textbf {\bibinfo {volume} {8}},\ \bibinfo {pages} {152} (\bibinfo {year}
  {2013})}\BibitemShut {NoStop}%
\bibitem [{\citenamefont {Fert}\ \emph {et~al.}(2017)\citenamefont {Fert},
  \citenamefont {Reyren},\ and\ \citenamefont {Cros}}]{fert2017magnetic}%
  \BibitemOpen
  \bibfield  {author} {\bibinfo {author} {\bibfnamefont {A.}~\bibnamefont
  {Fert}}, \bibinfo {author} {\bibfnamefont {N.}~\bibnamefont {Reyren}}, \ and\
  \bibinfo {author} {\bibfnamefont {V.}~\bibnamefont {Cros}},\ }\href@noop {}
  {\bibfield  {journal} {\bibinfo  {journal} {Nature Reviews Materials}\
  }\textbf {\bibinfo {volume} {2}},\ \bibinfo {pages} {1} (\bibinfo {year}
  {2017})}\BibitemShut {NoStop}%
\bibitem [{\citenamefont {Song}\ \emph {et~al.}(2020)\citenamefont {Song},
  \citenamefont {Jeong}, \citenamefont {Pan}, \citenamefont {Zhang},
  \citenamefont {Xia}, \citenamefont {Cha}, \citenamefont {Park}, \citenamefont
  {Kim}, \citenamefont {Finizio}, \citenamefont {Raabe} \emph
  {et~al.}}]{song2020skyrmion}%
  \BibitemOpen
  \bibfield  {author} {\bibinfo {author} {\bibfnamefont {K.~M.}\ \bibnamefont
  {Song}}, \bibinfo {author} {\bibfnamefont {J.-S.}\ \bibnamefont {Jeong}},
  \bibinfo {author} {\bibfnamefont {B.}~\bibnamefont {Pan}}, \bibinfo {author}
  {\bibfnamefont {X.}~\bibnamefont {Zhang}}, \bibinfo {author} {\bibfnamefont
  {J.}~\bibnamefont {Xia}}, \bibinfo {author} {\bibfnamefont {S.}~\bibnamefont
  {Cha}}, \bibinfo {author} {\bibfnamefont {T.-E.}\ \bibnamefont {Park}},
  \bibinfo {author} {\bibfnamefont {K.}~\bibnamefont {Kim}}, \bibinfo {author}
  {\bibfnamefont {S.}~\bibnamefont {Finizio}}, \bibinfo {author} {\bibfnamefont
  {J.}~\bibnamefont {Raabe}},  \emph {et~al.},\ }\href@noop {} {\bibfield
  {journal} {\bibinfo  {journal} {Nature Electronics}\ }\textbf {\bibinfo
  {volume} {3}},\ \bibinfo {pages} {148} (\bibinfo {year} {2020})}\BibitemShut
  {NoStop}%
\bibitem [{\citenamefont {Jonietz}\ \emph {et~al.}(2010)\citenamefont
  {Jonietz}, \citenamefont {M{\"u}hlbauer}, \citenamefont {Pfleiderer},
  \citenamefont {Neubauer}, \citenamefont {M{\"u}nzer}, \citenamefont {Bauer},
  \citenamefont {Adams}, \citenamefont {Georgii}, \citenamefont {B{\"o}ni},
  \citenamefont {Duine} \emph {et~al.}}]{jonietz2010spin}%
  \BibitemOpen
  \bibfield  {author} {\bibinfo {author} {\bibfnamefont {F.}~\bibnamefont
  {Jonietz}}, \bibinfo {author} {\bibfnamefont {S.}~\bibnamefont
  {M{\"u}hlbauer}}, \bibinfo {author} {\bibfnamefont {C.}~\bibnamefont
  {Pfleiderer}}, \bibinfo {author} {\bibfnamefont {A.}~\bibnamefont
  {Neubauer}}, \bibinfo {author} {\bibfnamefont {W.}~\bibnamefont
  {M{\"u}nzer}}, \bibinfo {author} {\bibfnamefont {A.}~\bibnamefont {Bauer}},
  \bibinfo {author} {\bibfnamefont {T.}~\bibnamefont {Adams}}, \bibinfo
  {author} {\bibfnamefont {R.}~\bibnamefont {Georgii}}, \bibinfo {author}
  {\bibfnamefont {P.}~\bibnamefont {B{\"o}ni}}, \bibinfo {author}
  {\bibfnamefont {R.~A.}\ \bibnamefont {Duine}},  \emph {et~al.},\ }\href@noop
  {} {\bibfield  {journal} {\bibinfo  {journal} {Science}\ }\textbf {\bibinfo
  {volume} {330}},\ \bibinfo {pages} {1648} (\bibinfo {year}
  {2010})}\BibitemShut {NoStop}%
\bibitem [{\citenamefont {M{\"u}hlbauer}\ \emph {et~al.}(2009)\citenamefont
  {M{\"u}hlbauer}, \citenamefont {Binz}, \citenamefont {Jonietz}, \citenamefont
  {Pfleiderer}, \citenamefont {Rosch}, \citenamefont {Neubauer}, \citenamefont
  {Georgii},\ and\ \citenamefont {B{\"o}ni}}]{muhlbauer2009skyrmion}%
  \BibitemOpen
  \bibfield  {author} {\bibinfo {author} {\bibfnamefont {S.}~\bibnamefont
  {M{\"u}hlbauer}}, \bibinfo {author} {\bibfnamefont {B.}~\bibnamefont {Binz}},
  \bibinfo {author} {\bibfnamefont {F.}~\bibnamefont {Jonietz}}, \bibinfo
  {author} {\bibfnamefont {C.}~\bibnamefont {Pfleiderer}}, \bibinfo {author}
  {\bibfnamefont {A.}~\bibnamefont {Rosch}}, \bibinfo {author} {\bibfnamefont
  {A.}~\bibnamefont {Neubauer}}, \bibinfo {author} {\bibfnamefont
  {R.}~\bibnamefont {Georgii}}, \ and\ \bibinfo {author} {\bibfnamefont
  {P.}~\bibnamefont {B{\"o}ni}},\ }\href@noop {} {\bibfield  {journal}
  {\bibinfo  {journal} {Science}\ }\textbf {\bibinfo {volume} {323}},\ \bibinfo
  {pages} {915} (\bibinfo {year} {2009})}\BibitemShut {NoStop}%
\bibitem [{\citenamefont {Yu}\ \emph {et~al.}(2011)\citenamefont {Yu},
  \citenamefont {Kanazawa}, \citenamefont {Onose}, \citenamefont {Kimoto},
  \citenamefont {Zhang}, \citenamefont {Ishiwata}, \citenamefont {Matsui},\
  and\ \citenamefont {Tokura}}]{yu2011near}%
  \BibitemOpen
  \bibfield  {author} {\bibinfo {author} {\bibfnamefont {X.}~\bibnamefont
  {Yu}}, \bibinfo {author} {\bibfnamefont {N.}~\bibnamefont {Kanazawa}},
  \bibinfo {author} {\bibfnamefont {Y.}~\bibnamefont {Onose}}, \bibinfo
  {author} {\bibfnamefont {K.}~\bibnamefont {Kimoto}}, \bibinfo {author}
  {\bibfnamefont {W.}~\bibnamefont {Zhang}}, \bibinfo {author} {\bibfnamefont
  {S.}~\bibnamefont {Ishiwata}}, \bibinfo {author} {\bibfnamefont
  {Y.}~\bibnamefont {Matsui}}, \ and\ \bibinfo {author} {\bibfnamefont
  {Y.}~\bibnamefont {Tokura}},\ }\href@noop {} {\bibfield  {journal} {\bibinfo
  {journal} {Nature materials}\ }\textbf {\bibinfo {volume} {10}},\ \bibinfo
  {pages} {106} (\bibinfo {year} {2011})}\BibitemShut {NoStop}%
\bibitem [{\citenamefont {Seki}\ \emph {et~al.}(2012)\citenamefont {Seki},
  \citenamefont {Yu}, \citenamefont {Ishiwata},\ and\ \citenamefont
  {Tokura}}]{seki2012observation}%
  \BibitemOpen
  \bibfield  {author} {\bibinfo {author} {\bibfnamefont {S.}~\bibnamefont
  {Seki}}, \bibinfo {author} {\bibfnamefont {X.}~\bibnamefont {Yu}}, \bibinfo
  {author} {\bibfnamefont {S.}~\bibnamefont {Ishiwata}}, \ and\ \bibinfo
  {author} {\bibfnamefont {Y.}~\bibnamefont {Tokura}},\ }\href@noop {}
  {\bibfield  {journal} {\bibinfo  {journal} {Science}\ }\textbf {\bibinfo
  {volume} {336}},\ \bibinfo {pages} {198} (\bibinfo {year}
  {2012})}\BibitemShut {NoStop}%
\bibitem [{\citenamefont {Adams}\ \emph {et~al.}(2012)\citenamefont {Adams},
  \citenamefont {Chacon}, \citenamefont {Wagner}, \citenamefont {Bauer},
  \citenamefont {Brandl}, \citenamefont {Pedersen}, \citenamefont {Berger},
  \citenamefont {Lemmens},\ and\ \citenamefont {Pfleiderer}}]{adams2012long}%
  \BibitemOpen
  \bibfield  {author} {\bibinfo {author} {\bibfnamefont {T.}~\bibnamefont
  {Adams}}, \bibinfo {author} {\bibfnamefont {A.}~\bibnamefont {Chacon}},
  \bibinfo {author} {\bibfnamefont {M.}~\bibnamefont {Wagner}}, \bibinfo
  {author} {\bibfnamefont {A.}~\bibnamefont {Bauer}}, \bibinfo {author}
  {\bibfnamefont {G.}~\bibnamefont {Brandl}}, \bibinfo {author} {\bibfnamefont
  {B.}~\bibnamefont {Pedersen}}, \bibinfo {author} {\bibfnamefont
  {H.}~\bibnamefont {Berger}}, \bibinfo {author} {\bibfnamefont
  {P.}~\bibnamefont {Lemmens}}, \ and\ \bibinfo {author} {\bibfnamefont
  {C.}~\bibnamefont {Pfleiderer}},\ }\href@noop {} {\bibfield  {journal}
  {\bibinfo  {journal} {Physical review letters}\ }\textbf {\bibinfo {volume}
  {108}},\ \bibinfo {pages} {237204} (\bibinfo {year} {2012})}\BibitemShut
  {NoStop}%
\bibitem [{\citenamefont
  {Dzyaloshinsky}(1958)}]{dzyaloshinsky1958thermodynamic}%
  \BibitemOpen
  \bibfield  {author} {\bibinfo {author} {\bibfnamefont {I.}~\bibnamefont
  {Dzyaloshinsky}},\ }\href@noop {} {\bibfield  {journal} {\bibinfo  {journal}
  {Journal of Physics and Chemistry of Solids}\ }\textbf {\bibinfo {volume}
  {4}},\ \bibinfo {pages} {241} (\bibinfo {year} {1958})}\BibitemShut {NoStop}%
\bibitem [{\citenamefont {Moriya}(1960)}]{moriya1960new}%
  \BibitemOpen
  \bibfield  {author} {\bibinfo {author} {\bibfnamefont {T.}~\bibnamefont
  {Moriya}},\ }\href@noop {} {\bibfield  {journal} {\bibinfo  {journal}
  {Physical Review Letters}\ }\textbf {\bibinfo {volume} {4}},\ \bibinfo
  {pages} {228} (\bibinfo {year} {1960})}\BibitemShut {NoStop}%
\bibitem [{\citenamefont {Okubo}\ \emph {et~al.}(2012)\citenamefont {Okubo},
  \citenamefont {Chung},\ and\ \citenamefont {Kawamura}}]{okubo2012multiple}%
  \BibitemOpen
  \bibfield  {author} {\bibinfo {author} {\bibfnamefont {T.}~\bibnamefont
  {Okubo}}, \bibinfo {author} {\bibfnamefont {S.}~\bibnamefont {Chung}}, \ and\
  \bibinfo {author} {\bibfnamefont {H.}~\bibnamefont {Kawamura}},\ }\href@noop
  {} {\bibfield  {journal} {\bibinfo  {journal} {Phys. Rev. Lett.}\ }\textbf
  {\bibinfo {volume} {108}},\ \bibinfo {pages} {017206} (\bibinfo {year}
  {2012})}\BibitemShut {NoStop}%
\bibitem [{\citenamefont {Leonov}\ and\ \citenamefont
  {Mostovoy}(2015)}]{leonov2015multiply}%
  \BibitemOpen
  \bibfield  {author} {\bibinfo {author} {\bibfnamefont {A.~O.}\ \bibnamefont
  {Leonov}}\ and\ \bibinfo {author} {\bibfnamefont {M.}~\bibnamefont
  {Mostovoy}},\ }\href@noop {} {\bibfield  {journal} {\bibinfo  {journal}
  {Nature Communications}\ }\textbf {\bibinfo {volume} {6}},\ \bibinfo {pages}
  {8275} (\bibinfo {year} {2015})}\BibitemShut {NoStop}%
\bibitem [{\citenamefont {Lin}\ and\ \citenamefont
  {Hayami}(2016)}]{lin2016ginzburg}%
  \BibitemOpen
  \bibfield  {author} {\bibinfo {author} {\bibfnamefont {S.-Z.}\ \bibnamefont
  {Lin}}\ and\ \bibinfo {author} {\bibfnamefont {S.}~\bibnamefont {Hayami}},\
  }\href@noop {} {\bibfield  {journal} {\bibinfo  {journal} {Physical Review
  B}\ }\textbf {\bibinfo {volume} {93}},\ \bibinfo {pages} {064430} (\bibinfo
  {year} {2016})}\BibitemShut {NoStop}%
\bibitem [{\citenamefont {Heinze}\ \emph {et~al.}(2011)\citenamefont {Heinze},
  \citenamefont {Von~Bergmann}, \citenamefont {Menzel}, \citenamefont {Brede},
  \citenamefont {Kubetzka}, \citenamefont {Wiesendanger}, \citenamefont
  {Bihlmayer},\ and\ \citenamefont {Bl{\"u}gel}}]{heinze2011spontaneous}%
  \BibitemOpen
  \bibfield  {author} {\bibinfo {author} {\bibfnamefont {S.}~\bibnamefont
  {Heinze}}, \bibinfo {author} {\bibfnamefont {K.}~\bibnamefont
  {Von~Bergmann}}, \bibinfo {author} {\bibfnamefont {M.}~\bibnamefont
  {Menzel}}, \bibinfo {author} {\bibfnamefont {J.}~\bibnamefont {Brede}},
  \bibinfo {author} {\bibfnamefont {A.}~\bibnamefont {Kubetzka}}, \bibinfo
  {author} {\bibfnamefont {R.}~\bibnamefont {Wiesendanger}}, \bibinfo {author}
  {\bibfnamefont {G.}~\bibnamefont {Bihlmayer}}, \ and\ \bibinfo {author}
  {\bibfnamefont {S.}~\bibnamefont {Bl{\"u}gel}},\ }\href@noop {} {\bibfield
  {journal} {\bibinfo  {journal} {Nature Physics}\ }\textbf {\bibinfo {volume}
  {7}},\ \bibinfo {pages} {713} (\bibinfo {year} {2011})}\BibitemShut {NoStop}%
\bibitem [{\citenamefont {Batista}\ \emph {et~al.}(2016)\citenamefont
  {Batista}, \citenamefont {Lin}, \citenamefont {Hayami},\ and\ \citenamefont
  {Kamiya}}]{batista2016frustration}%
  \BibitemOpen
  \bibfield  {author} {\bibinfo {author} {\bibfnamefont {C.~D.}\ \bibnamefont
  {Batista}}, \bibinfo {author} {\bibfnamefont {S.-Z.}\ \bibnamefont {Lin}},
  \bibinfo {author} {\bibfnamefont {S.}~\bibnamefont {Hayami}}, \ and\ \bibinfo
  {author} {\bibfnamefont {Y.}~\bibnamefont {Kamiya}},\ }\href@noop {}
  {\bibfield  {journal} {\bibinfo  {journal} {Reports on Progress in Physics}\
  }\textbf {\bibinfo {volume} {79}},\ \bibinfo {pages} {084504} (\bibinfo
  {year} {2016})}\BibitemShut {NoStop}%
\bibitem [{\citenamefont {Ozawa}\ \emph {et~al.}(2017)\citenamefont {Ozawa},
  \citenamefont {Hayami},\ and\ \citenamefont {Motome}}]{ozawa2017zero}%
  \BibitemOpen
  \bibfield  {author} {\bibinfo {author} {\bibfnamefont {R.}~\bibnamefont
  {Ozawa}}, \bibinfo {author} {\bibfnamefont {S.}~\bibnamefont {Hayami}}, \
  and\ \bibinfo {author} {\bibfnamefont {Y.}~\bibnamefont {Motome}},\
  }\href@noop {} {\bibfield  {journal} {\bibinfo  {journal} {Physical Review
  Letters}\ }\textbf {\bibinfo {volume} {118}},\ \bibinfo {pages} {147205}
  (\bibinfo {year} {2017})}\BibitemShut {NoStop}%
\bibitem [{\citenamefont {Hayami}\ \emph {et~al.}(2017)\citenamefont {Hayami},
  \citenamefont {Ozawa},\ and\ \citenamefont {Motome}}]{hayami2017effective}%
  \BibitemOpen
  \bibfield  {author} {\bibinfo {author} {\bibfnamefont {S.}~\bibnamefont
  {Hayami}}, \bibinfo {author} {\bibfnamefont {R.}~\bibnamefont {Ozawa}}, \
  and\ \bibinfo {author} {\bibfnamefont {Y.}~\bibnamefont {Motome}},\
  }\href@noop {} {\bibfield  {journal} {\bibinfo  {journal} {Physical Review
  B}\ }\textbf {\bibinfo {volume} {95}},\ \bibinfo {pages} {224424} (\bibinfo
  {year} {2017})}\BibitemShut {NoStop}%
\bibitem [{\citenamefont {Hou}\ \emph {et~al.}(2017)\citenamefont {Hou},
  \citenamefont {Ren}, \citenamefont {Ding}, \citenamefont {Xu}, \citenamefont
  {Wang}, \citenamefont {Yang}, \citenamefont {Zhang}, \citenamefont {Zhang},
  \citenamefont {Liu}, \citenamefont {Xu} \emph {et~al.}}]{hou2017observation}%
  \BibitemOpen
  \bibfield  {author} {\bibinfo {author} {\bibfnamefont {Z.}~\bibnamefont
  {Hou}}, \bibinfo {author} {\bibfnamefont {W.}~\bibnamefont {Ren}}, \bibinfo
  {author} {\bibfnamefont {B.}~\bibnamefont {Ding}}, \bibinfo {author}
  {\bibfnamefont {G.}~\bibnamefont {Xu}}, \bibinfo {author} {\bibfnamefont
  {Y.}~\bibnamefont {Wang}}, \bibinfo {author} {\bibfnamefont {B.}~\bibnamefont
  {Yang}}, \bibinfo {author} {\bibfnamefont {Q.}~\bibnamefont {Zhang}},
  \bibinfo {author} {\bibfnamefont {Y.}~\bibnamefont {Zhang}}, \bibinfo
  {author} {\bibfnamefont {E.}~\bibnamefont {Liu}}, \bibinfo {author}
  {\bibfnamefont {F.}~\bibnamefont {Xu}},  \emph {et~al.},\ }\href@noop {}
  {\bibfield  {journal} {\bibinfo  {journal} {Advanced Materials}\ }\textbf
  {\bibinfo {volume} {29}},\ \bibinfo {pages} {1701144} (\bibinfo {year}
  {2017})}\BibitemShut {NoStop}%
\bibitem [{\citenamefont {Xiao}\ \emph {et~al.}(2020)\citenamefont {Xiao},
  \citenamefont {Morvan}, \citenamefont {He}, \citenamefont {Wang},
  \citenamefont {Luo}, \citenamefont {Jiao}, \citenamefont {Xia}, \citenamefont
  {Zhao},\ and\ \citenamefont {Liu}}]{xiao2020spin}%
  \BibitemOpen
  \bibfield  {author} {\bibinfo {author} {\bibfnamefont {Y.}~\bibnamefont
  {Xiao}}, \bibinfo {author} {\bibfnamefont {F.}~\bibnamefont {Morvan}},
  \bibinfo {author} {\bibfnamefont {A.}~\bibnamefont {He}}, \bibinfo {author}
  {\bibfnamefont {M.}~\bibnamefont {Wang}}, \bibinfo {author} {\bibfnamefont
  {H.}~\bibnamefont {Luo}}, \bibinfo {author} {\bibfnamefont {R.}~\bibnamefont
  {Jiao}}, \bibinfo {author} {\bibfnamefont {W.}~\bibnamefont {Xia}}, \bibinfo
  {author} {\bibfnamefont {G.}~\bibnamefont {Zhao}}, \ and\ \bibinfo {author}
  {\bibfnamefont {J.}~\bibnamefont {Liu}},\ }\href@noop {} {\bibfield
  {journal} {\bibinfo  {journal} {Applied Physics Letters}\ }\textbf {\bibinfo
  {volume} {117}},\ \bibinfo {pages} {132402} (\bibinfo {year}
  {2020})}\BibitemShut {NoStop}%
\bibitem [{\citenamefont {He}\ \emph {et~al.}(2022)\citenamefont {He},
  \citenamefont {Helm}, \citenamefont {Soldatov}, \citenamefont {Schneider},
  \citenamefont {Pohl}, \citenamefont {Srivastava}, \citenamefont {Sharma},
  \citenamefont {Kroder}, \citenamefont {Schnelle}, \citenamefont {Schaefer}
  \emph {et~al.}}]{he2022nanoscale}%
  \BibitemOpen
  \bibfield  {author} {\bibinfo {author} {\bibfnamefont {Y.}~\bibnamefont
  {He}}, \bibinfo {author} {\bibfnamefont {T.}~\bibnamefont {Helm}}, \bibinfo
  {author} {\bibfnamefont {I.}~\bibnamefont {Soldatov}}, \bibinfo {author}
  {\bibfnamefont {S.}~\bibnamefont {Schneider}}, \bibinfo {author}
  {\bibfnamefont {D.}~\bibnamefont {Pohl}}, \bibinfo {author} {\bibfnamefont
  {A.~K.}\ \bibnamefont {Srivastava}}, \bibinfo {author} {\bibfnamefont
  {A.~K.}\ \bibnamefont {Sharma}}, \bibinfo {author} {\bibfnamefont
  {J.}~\bibnamefont {Kroder}}, \bibinfo {author} {\bibfnamefont
  {W.}~\bibnamefont {Schnelle}}, \bibinfo {author} {\bibfnamefont
  {R.}~\bibnamefont {Schaefer}},  \emph {et~al.},\ }\href@noop {} {\bibfield
  {journal} {\bibinfo  {journal} {Physical Review B}\ }\textbf {\bibinfo
  {volume} {105}},\ \bibinfo {pages} {064426} (\bibinfo {year}
  {2022})}\BibitemShut {NoStop}%
\bibitem [{\citenamefont {Kurumaji}\ \emph {et~al.}(2019)\citenamefont
  {Kurumaji}, \citenamefont {Nakajima}, \citenamefont {Hirschberger},
  \citenamefont {Kikkawa}, \citenamefont {Yamasaki}, \citenamefont {Sagayama},
  \citenamefont {Nakao}, \citenamefont {Taguchi}, \citenamefont {Arima},\ and\
  \citenamefont {Tokura}}]{kurumaji2019skyrmion}%
  \BibitemOpen
  \bibfield  {author} {\bibinfo {author} {\bibfnamefont {T.}~\bibnamefont
  {Kurumaji}}, \bibinfo {author} {\bibfnamefont {T.}~\bibnamefont {Nakajima}},
  \bibinfo {author} {\bibfnamefont {M.}~\bibnamefont {Hirschberger}}, \bibinfo
  {author} {\bibfnamefont {A.}~\bibnamefont {Kikkawa}}, \bibinfo {author}
  {\bibfnamefont {Y.}~\bibnamefont {Yamasaki}}, \bibinfo {author}
  {\bibfnamefont {H.}~\bibnamefont {Sagayama}}, \bibinfo {author}
  {\bibfnamefont {H.}~\bibnamefont {Nakao}}, \bibinfo {author} {\bibfnamefont
  {Y.}~\bibnamefont {Taguchi}}, \bibinfo {author} {\bibfnamefont {T.-h.}\
  \bibnamefont {Arima}}, \ and\ \bibinfo {author} {\bibfnamefont
  {Y.}~\bibnamefont {Tokura}},\ }\href@noop {} {\bibfield  {journal} {\bibinfo
  {journal} {Science}\ }\textbf {\bibinfo {volume} {365}},\ \bibinfo {pages}
  {914} (\bibinfo {year} {2019})}\BibitemShut {NoStop}%
\bibitem [{\citenamefont {Hirschberger}\ \emph {et~al.}(2020)\citenamefont
  {Hirschberger}, \citenamefont {Spitz}, \citenamefont {Nomoto}, \citenamefont
  {Kurumaji}, \citenamefont {Gao}, \citenamefont {Masell}, \citenamefont
  {Nakajima}, \citenamefont {Kikkawa}, \citenamefont {Yamasaki}, \citenamefont
  {Sagayama} \emph {et~al.}}]{hirschberger2020topological}%
  \BibitemOpen
  \bibfield  {author} {\bibinfo {author} {\bibfnamefont {M.}~\bibnamefont
  {Hirschberger}}, \bibinfo {author} {\bibfnamefont {L.}~\bibnamefont {Spitz}},
  \bibinfo {author} {\bibfnamefont {T.}~\bibnamefont {Nomoto}}, \bibinfo
  {author} {\bibfnamefont {T.}~\bibnamefont {Kurumaji}}, \bibinfo {author}
  {\bibfnamefont {S.}~\bibnamefont {Gao}}, \bibinfo {author} {\bibfnamefont
  {J.}~\bibnamefont {Masell}}, \bibinfo {author} {\bibfnamefont
  {T.}~\bibnamefont {Nakajima}}, \bibinfo {author} {\bibfnamefont
  {A.}~\bibnamefont {Kikkawa}}, \bibinfo {author} {\bibfnamefont
  {Y.}~\bibnamefont {Yamasaki}}, \bibinfo {author} {\bibfnamefont
  {H.}~\bibnamefont {Sagayama}},  \emph {et~al.},\ }\href@noop {} {\bibfield
  {journal} {\bibinfo  {journal} {Physical Review Letters}\ }\textbf {\bibinfo
  {volume} {125}},\ \bibinfo {pages} {076602} (\bibinfo {year}
  {2020})}\BibitemShut {NoStop}%
\bibitem [{\citenamefont {Hirschberger}\ \emph {et~al.}(2019)\citenamefont
  {Hirschberger}, \citenamefont {Nakajima}, \citenamefont {Gao}, \citenamefont
  {Peng}, \citenamefont {Kikkawa}, \citenamefont {Kurumaji}, \citenamefont
  {Kriener}, \citenamefont {Yamasaki}, \citenamefont {Sagayama}, \citenamefont
  {Nakao} \emph {et~al.}}]{hirschberger2019skyrmion}%
  \BibitemOpen
  \bibfield  {author} {\bibinfo {author} {\bibfnamefont {M.}~\bibnamefont
  {Hirschberger}}, \bibinfo {author} {\bibfnamefont {T.}~\bibnamefont
  {Nakajima}}, \bibinfo {author} {\bibfnamefont {S.}~\bibnamefont {Gao}},
  \bibinfo {author} {\bibfnamefont {L.}~\bibnamefont {Peng}}, \bibinfo {author}
  {\bibfnamefont {A.}~\bibnamefont {Kikkawa}}, \bibinfo {author} {\bibfnamefont
  {T.}~\bibnamefont {Kurumaji}}, \bibinfo {author} {\bibfnamefont
  {M.}~\bibnamefont {Kriener}}, \bibinfo {author} {\bibfnamefont
  {Y.}~\bibnamefont {Yamasaki}}, \bibinfo {author} {\bibfnamefont
  {H.}~\bibnamefont {Sagayama}}, \bibinfo {author} {\bibfnamefont
  {H.}~\bibnamefont {Nakao}},  \emph {et~al.},\ }\href@noop {} {\bibfield
  {journal} {\bibinfo  {journal} {Nature communications}\ }\textbf {\bibinfo
  {volume} {10}},\ \bibinfo {pages} {1} (\bibinfo {year} {2019})}\BibitemShut
  {NoStop}%
\bibitem [{\citenamefont {Khanh}\ \emph {et~al.}(2020)\citenamefont {Khanh},
  \citenamefont {Nakajima}, \citenamefont {Yu}, \citenamefont {Gao},
  \citenamefont {Shibata}, \citenamefont {Hirschberger}, \citenamefont
  {Yamasaki}, \citenamefont {Sagayama}, \citenamefont {Nakao}, \citenamefont
  {Peng} \emph {et~al.}}]{khanh2020nanometric}%
  \BibitemOpen
  \bibfield  {author} {\bibinfo {author} {\bibfnamefont {N.~D.}\ \bibnamefont
  {Khanh}}, \bibinfo {author} {\bibfnamefont {T.}~\bibnamefont {Nakajima}},
  \bibinfo {author} {\bibfnamefont {X.}~\bibnamefont {Yu}}, \bibinfo {author}
  {\bibfnamefont {S.}~\bibnamefont {Gao}}, \bibinfo {author} {\bibfnamefont
  {K.}~\bibnamefont {Shibata}}, \bibinfo {author} {\bibfnamefont
  {M.}~\bibnamefont {Hirschberger}}, \bibinfo {author} {\bibfnamefont
  {Y.}~\bibnamefont {Yamasaki}}, \bibinfo {author} {\bibfnamefont
  {H.}~\bibnamefont {Sagayama}}, \bibinfo {author} {\bibfnamefont
  {H.}~\bibnamefont {Nakao}}, \bibinfo {author} {\bibfnamefont
  {L.}~\bibnamefont {Peng}},  \emph {et~al.},\ }\href@noop {} {\bibfield
  {journal} {\bibinfo  {journal} {Nature Nanotechnology}\ }\textbf {\bibinfo
  {volume} {15}},\ \bibinfo {pages} {444} (\bibinfo {year} {2020})}\BibitemShut
  {NoStop}%
\bibitem [{\citenamefont {Yasui}\ \emph {et~al.}(2020)\citenamefont {Yasui},
  \citenamefont {Butler}, \citenamefont {Khanh}, \citenamefont {Hayami},
  \citenamefont {Nomoto}, \citenamefont {Hanaguri}, \citenamefont {Motome},
  \citenamefont {Arita}, \citenamefont {Arima}, \citenamefont {Tokura} \emph
  {et~al.}}]{yasui2020imaging}%
  \BibitemOpen
  \bibfield  {author} {\bibinfo {author} {\bibfnamefont {Y.}~\bibnamefont
  {Yasui}}, \bibinfo {author} {\bibfnamefont {C.~J.}\ \bibnamefont {Butler}},
  \bibinfo {author} {\bibfnamefont {N.~D.}\ \bibnamefont {Khanh}}, \bibinfo
  {author} {\bibfnamefont {S.}~\bibnamefont {Hayami}}, \bibinfo {author}
  {\bibfnamefont {T.}~\bibnamefont {Nomoto}}, \bibinfo {author} {\bibfnamefont
  {T.}~\bibnamefont {Hanaguri}}, \bibinfo {author} {\bibfnamefont
  {Y.}~\bibnamefont {Motome}}, \bibinfo {author} {\bibfnamefont
  {R.}~\bibnamefont {Arita}}, \bibinfo {author} {\bibfnamefont {T.-h.}\
  \bibnamefont {Arima}}, \bibinfo {author} {\bibfnamefont {Y.}~\bibnamefont
  {Tokura}},  \emph {et~al.},\ }\href@noop {} {\bibfield  {journal} {\bibinfo
  {journal} {Nature communications}\ }\textbf {\bibinfo {volume} {11}},\
  \bibinfo {pages} {1} (\bibinfo {year} {2020})}\BibitemShut {NoStop}%
\bibitem [{\citenamefont {Ishiwata}\ \emph {et~al.}(2011)\citenamefont
  {Ishiwata}, \citenamefont {Tokunaga}, \citenamefont {Kaneko}, \citenamefont
  {Okuyama}, \citenamefont {Tokunaga}, \citenamefont {Wakimoto}, \citenamefont
  {Kakurai}, \citenamefont {Arima}, \citenamefont {Taguchi},\ and\
  \citenamefont {Tokura}}]{ishiwata2011versatile}%
  \BibitemOpen
  \bibfield  {author} {\bibinfo {author} {\bibfnamefont {S.}~\bibnamefont
  {Ishiwata}}, \bibinfo {author} {\bibfnamefont {M.}~\bibnamefont {Tokunaga}},
  \bibinfo {author} {\bibfnamefont {Y.}~\bibnamefont {Kaneko}}, \bibinfo
  {author} {\bibfnamefont {D.}~\bibnamefont {Okuyama}}, \bibinfo {author}
  {\bibfnamefont {Y.}~\bibnamefont {Tokunaga}}, \bibinfo {author}
  {\bibfnamefont {S.}~\bibnamefont {Wakimoto}}, \bibinfo {author}
  {\bibfnamefont {K.}~\bibnamefont {Kakurai}}, \bibinfo {author} {\bibfnamefont
  {T.}~\bibnamefont {Arima}}, \bibinfo {author} {\bibfnamefont
  {Y.}~\bibnamefont {Taguchi}}, \ and\ \bibinfo {author} {\bibfnamefont
  {Y.}~\bibnamefont {Tokura}},\ }\href@noop {} {\bibfield  {journal} {\bibinfo
  {journal} {Physical Review B}\ }\textbf {\bibinfo {volume} {84}},\ \bibinfo
  {pages} {054427} (\bibinfo {year} {2011})}\BibitemShut {NoStop}%
\bibitem [{\citenamefont {Ishiwata}\ \emph {et~al.}(2020)\citenamefont
  {Ishiwata}, \citenamefont {Nakajima}, \citenamefont {Kim}, \citenamefont
  {Inosov}, \citenamefont {Kanazawa}, \citenamefont {White}, \citenamefont
  {Gavilano}, \citenamefont {Georgii}, \citenamefont {Seemann}, \citenamefont
  {Brandl} \emph {et~al.}}]{ishiwata2020emergent}%
  \BibitemOpen
  \bibfield  {author} {\bibinfo {author} {\bibfnamefont {S.}~\bibnamefont
  {Ishiwata}}, \bibinfo {author} {\bibfnamefont {T.}~\bibnamefont {Nakajima}},
  \bibinfo {author} {\bibfnamefont {J.-H.}\ \bibnamefont {Kim}}, \bibinfo
  {author} {\bibfnamefont {D.}~\bibnamefont {Inosov}}, \bibinfo {author}
  {\bibfnamefont {N.}~\bibnamefont {Kanazawa}}, \bibinfo {author}
  {\bibfnamefont {J.}~\bibnamefont {White}}, \bibinfo {author} {\bibfnamefont
  {J.}~\bibnamefont {Gavilano}}, \bibinfo {author} {\bibfnamefont
  {R.}~\bibnamefont {Georgii}}, \bibinfo {author} {\bibfnamefont
  {K.}~\bibnamefont {Seemann}}, \bibinfo {author} {\bibfnamefont
  {G.}~\bibnamefont {Brandl}},  \emph {et~al.},\ }\href@noop {} {\bibfield
  {journal} {\bibinfo  {journal} {Physical Review B}\ }\textbf {\bibinfo
  {volume} {101}},\ \bibinfo {pages} {134406} (\bibinfo {year}
  {2020})}\BibitemShut {NoStop}%
\bibitem [{\citenamefont {Nomoto}\ \emph {et~al.}(2020)\citenamefont {Nomoto},
  \citenamefont {Koretsune},\ and\ \citenamefont
  {Arita}}]{Nomoto2020formation}%
  \BibitemOpen
  \bibfield  {author} {\bibinfo {author} {\bibfnamefont {T.}~\bibnamefont
  {Nomoto}}, \bibinfo {author} {\bibfnamefont {T.}~\bibnamefont {Koretsune}}, \
  and\ \bibinfo {author} {\bibfnamefont {R.}~\bibnamefont {Arita}},\
  }\href@noop {} {\bibfield  {journal} {\bibinfo  {journal} {Phys. Rev. Lett.}\
  }\textbf {\bibinfo {volume} {125}},\ \bibinfo {pages} {117204} (\bibinfo
  {year} {2020})}\BibitemShut {NoStop}%
\bibitem [{\citenamefont {Hayami}\ and\ \citenamefont
  {Motome}(2020)}]{hayami2020square}%
  \BibitemOpen
  \bibfield  {author} {\bibinfo {author} {\bibfnamefont {S.}~\bibnamefont
  {Hayami}}\ and\ \bibinfo {author} {\bibfnamefont {Y.}~\bibnamefont
  {Motome}},\ }\href@noop {} {\bibfield  {journal} {\bibinfo  {journal} {arXiv
  preprint arXiv:2010.14671}\ } (\bibinfo {year} {2020})}\BibitemShut {NoStop}%
\bibitem [{\citenamefont {Wang}\ \emph {et~al.}(2021)\citenamefont {Wang},
  \citenamefont {Su}, \citenamefont {Lin},\ and\ \citenamefont
  {Batista}}]{wang2021meron}%
  \BibitemOpen
  \bibfield  {author} {\bibinfo {author} {\bibfnamefont {Z.}~\bibnamefont
  {Wang}}, \bibinfo {author} {\bibfnamefont {Y.}~\bibnamefont {Su}}, \bibinfo
  {author} {\bibfnamefont {S.-Z.}\ \bibnamefont {Lin}}, \ and\ \bibinfo
  {author} {\bibfnamefont {C.~D.}\ \bibnamefont {Batista}},\ }\href@noop {}
  {\bibfield  {journal} {\bibinfo  {journal} {Physical Review B}\ }\textbf
  {\bibinfo {volume} {103}},\ \bibinfo {pages} {104408} (\bibinfo {year}
  {2021})}\BibitemShut {NoStop}%
\bibitem [{\citenamefont {Stavinoha}\ \emph {et~al.}(2018)\citenamefont
  {Stavinoha}, \citenamefont {Cooley}, \citenamefont {Minasian}, \citenamefont
  {McQueen}, \citenamefont {Kauzlarich}, \citenamefont {Huang},\ and\
  \citenamefont {Morosan}}]{stavinoha2018charge}%
  \BibitemOpen
  \bibfield  {author} {\bibinfo {author} {\bibfnamefont {M.}~\bibnamefont
  {Stavinoha}}, \bibinfo {author} {\bibfnamefont {J.~A.}\ \bibnamefont
  {Cooley}}, \bibinfo {author} {\bibfnamefont {S.~G.}\ \bibnamefont
  {Minasian}}, \bibinfo {author} {\bibfnamefont {T.~M.}\ \bibnamefont
  {McQueen}}, \bibinfo {author} {\bibfnamefont {S.~M.}\ \bibnamefont
  {Kauzlarich}}, \bibinfo {author} {\bibfnamefont {C.-L.}\ \bibnamefont
  {Huang}}, \ and\ \bibinfo {author} {\bibfnamefont {E.}~\bibnamefont
  {Morosan}},\ }\href@noop {} {\bibfield  {journal} {\bibinfo  {journal}
  {Physical Review B}\ }\textbf {\bibinfo {volume} {97}},\ \bibinfo {pages}
  {195146} (\bibinfo {year} {2018})}\BibitemShut {NoStop}%
\bibitem [{\citenamefont
  {Rodr{\'\i}guez-Carvajal}(1993)}]{rodriguez1993recent}%
  \BibitemOpen
  \bibfield  {author} {\bibinfo {author} {\bibfnamefont {J.}~\bibnamefont
  {Rodr{\'\i}guez-Carvajal}},\ }\href@noop {} {\bibfield  {journal} {\bibinfo
  {journal} {Physica B: Condensed Matter}\ }\textbf {\bibinfo {volume} {192}},\
  \bibinfo {pages} {55} (\bibinfo {year} {1993})}\BibitemShut {NoStop}%
\bibitem [{\citenamefont {Krogstad}\ \emph {et~al.}(2020)\citenamefont
  {Krogstad}, \citenamefont {Rosenkranz}, \citenamefont {Wozniak},
  \citenamefont {Jennings}, \citenamefont {Ruff}, \citenamefont {Vaughey},\
  and\ \citenamefont {Osborn}}]{krogstad2020reciprocal}%
  \BibitemOpen
  \bibfield  {author} {\bibinfo {author} {\bibfnamefont {M.~J.}\ \bibnamefont
  {Krogstad}}, \bibinfo {author} {\bibfnamefont {S.}~\bibnamefont
  {Rosenkranz}}, \bibinfo {author} {\bibfnamefont {J.~M.}\ \bibnamefont
  {Wozniak}}, \bibinfo {author} {\bibfnamefont {G.}~\bibnamefont {Jennings}},
  \bibinfo {author} {\bibfnamefont {J.~P.}\ \bibnamefont {Ruff}}, \bibinfo
  {author} {\bibfnamefont {J.~T.}\ \bibnamefont {Vaughey}}, \ and\ \bibinfo
  {author} {\bibfnamefont {R.}~\bibnamefont {Osborn}},\ }\href@noop {}
  {\bibfield  {journal} {\bibinfo  {journal} {Nature materials}\ }\textbf
  {\bibinfo {volume} {19}},\ \bibinfo {pages} {63} (\bibinfo {year}
  {2020})}\BibitemShut {NoStop}%
\bibitem [{\citenamefont {{\'S}laski}\ \emph {et~al.}(1984)\citenamefont
  {{\'S}laski}, \citenamefont {Szytu{\l}a}, \citenamefont {Leciejewicz},\ and\
  \citenamefont {Zygmunt}}]{slaski1984magnetic}%
  \BibitemOpen
  \bibfield  {author} {\bibinfo {author} {\bibfnamefont {M.}~\bibnamefont
  {{\'S}laski}}, \bibinfo {author} {\bibfnamefont {A.}~\bibnamefont
  {Szytu{\l}a}}, \bibinfo {author} {\bibfnamefont {J.}~\bibnamefont
  {Leciejewicz}}, \ and\ \bibinfo {author} {\bibfnamefont {A.}~\bibnamefont
  {Zygmunt}},\ }\href@noop {} {\bibfield  {journal} {\bibinfo  {journal}
  {Journal of magnetism and magnetic materials}\ }\textbf {\bibinfo {volume}
  {46}},\ \bibinfo {pages} {114} (\bibinfo {year} {1984})}\BibitemShut
  {NoStop}%
\bibitem [{\citenamefont {Tokura}\ and\ \citenamefont
  {Kanazawa}(2020)}]{tokura2020magnetic}%
  \BibitemOpen
  \bibfield  {author} {\bibinfo {author} {\bibfnamefont {Y.}~\bibnamefont
  {Tokura}}\ and\ \bibinfo {author} {\bibfnamefont {N.}~\bibnamefont
  {Kanazawa}},\ }\href@noop {} {\bibfield  {journal} {\bibinfo  {journal}
  {Chemical Reviews}\ } (\bibinfo {year} {2020})}\BibitemShut {NoStop}%
\bibitem [{\citenamefont {Materials}(2021)}]{SM}%
  \BibitemOpen
  \bibfield  {author} {\bibinfo {author} {\bibfnamefont {S.~S.}\ \bibnamefont
  {Materials}},\ }\href@noop {} {\  (\bibinfo {year} {2021})}\BibitemShut
  {NoStop}%
\bibitem [{\citenamefont {Choi}\ \emph {et~al.}(2001)\citenamefont {Choi},
  \citenamefont {Lynn}, \citenamefont {Lopez}, \citenamefont {Gammel},
  \citenamefont {Canfield},\ and\ \citenamefont {Bud'Ko}}]{choi2001direct}%
  \BibitemOpen
  \bibfield  {author} {\bibinfo {author} {\bibfnamefont {S.-M.}\ \bibnamefont
  {Choi}}, \bibinfo {author} {\bibfnamefont {J.}~\bibnamefont {Lynn}}, \bibinfo
  {author} {\bibfnamefont {D.}~\bibnamefont {Lopez}}, \bibinfo {author}
  {\bibfnamefont {P.}~\bibnamefont {Gammel}}, \bibinfo {author} {\bibfnamefont
  {P.}~\bibnamefont {Canfield}}, \ and\ \bibinfo {author} {\bibfnamefont
  {S.}~\bibnamefont {Bud'Ko}},\ }\href@noop {} {\bibfield  {journal} {\bibinfo
  {journal} {Physical review letters}\ }\textbf {\bibinfo {volume} {87}},\
  \bibinfo {pages} {107001} (\bibinfo {year} {2001})}\BibitemShut {NoStop}%
\bibitem [{\citenamefont {Takagi}\ \emph {et~al.}(2022)\citenamefont {Takagi},
  \citenamefont {Matsuyama}, \citenamefont {Ukleev}, \citenamefont {Yu},
  \citenamefont {White}, \citenamefont {Francoual}, \citenamefont {Mardegan},
  \citenamefont {Hayami}, \citenamefont {Saito}, \citenamefont {Kaneko} \emph
  {et~al.}}]{takagi2022square}%
  \BibitemOpen
  \bibfield  {author} {\bibinfo {author} {\bibfnamefont {R.}~\bibnamefont
  {Takagi}}, \bibinfo {author} {\bibfnamefont {N.}~\bibnamefont {Matsuyama}},
  \bibinfo {author} {\bibfnamefont {V.}~\bibnamefont {Ukleev}}, \bibinfo
  {author} {\bibfnamefont {L.}~\bibnamefont {Yu}}, \bibinfo {author}
  {\bibfnamefont {J.~S.}\ \bibnamefont {White}}, \bibinfo {author}
  {\bibfnamefont {S.}~\bibnamefont {Francoual}}, \bibinfo {author}
  {\bibfnamefont {J.~R.}\ \bibnamefont {Mardegan}}, \bibinfo {author}
  {\bibfnamefont {S.}~\bibnamefont {Hayami}}, \bibinfo {author} {\bibfnamefont
  {H.}~\bibnamefont {Saito}}, \bibinfo {author} {\bibfnamefont
  {K.}~\bibnamefont {Kaneko}},  \emph {et~al.},\ }\href@noop {} {\bibfield
  {journal} {\bibinfo  {journal} {Nature communications}\ }\textbf {\bibinfo
  {volume} {13}},\ \bibinfo {pages} {1} (\bibinfo {year} {2022})}\BibitemShut
  {NoStop}%
\bibitem [{\citenamefont {Ueda}\ \emph {et~al.}(2012)\citenamefont {Ueda},
  \citenamefont {Iguchi}, \citenamefont {Suzuki}, \citenamefont {Ishiwata},
  \citenamefont {Taguchi},\ and\ \citenamefont {Tokura}}]{ueda2012topological}%
  \BibitemOpen
  \bibfield  {author} {\bibinfo {author} {\bibfnamefont {K.}~\bibnamefont
  {Ueda}}, \bibinfo {author} {\bibfnamefont {S.}~\bibnamefont {Iguchi}},
  \bibinfo {author} {\bibfnamefont {T.}~\bibnamefont {Suzuki}}, \bibinfo
  {author} {\bibfnamefont {S.}~\bibnamefont {Ishiwata}}, \bibinfo {author}
  {\bibfnamefont {Y.}~\bibnamefont {Taguchi}}, \ and\ \bibinfo {author}
  {\bibfnamefont {Y.}~\bibnamefont {Tokura}},\ }\href@noop {} {\bibfield
  {journal} {\bibinfo  {journal} {Physical review letters}\ }\textbf {\bibinfo
  {volume} {108}},\ \bibinfo {pages} {156601} (\bibinfo {year}
  {2012})}\BibitemShut {NoStop}%
\bibitem [{\citenamefont {Nagaosa}\ \emph {et~al.}(2010)\citenamefont
  {Nagaosa}, \citenamefont {Sinova}, \citenamefont {Onoda}, \citenamefont
  {MacDonald},\ and\ \citenamefont {Ong}}]{nagaosa2010anomalous}%
  \BibitemOpen
  \bibfield  {author} {\bibinfo {author} {\bibfnamefont {N.}~\bibnamefont
  {Nagaosa}}, \bibinfo {author} {\bibfnamefont {J.}~\bibnamefont {Sinova}},
  \bibinfo {author} {\bibfnamefont {S.}~\bibnamefont {Onoda}}, \bibinfo
  {author} {\bibfnamefont {A.~H.}\ \bibnamefont {MacDonald}}, \ and\ \bibinfo
  {author} {\bibfnamefont {N.~P.}\ \bibnamefont {Ong}},\ }\href@noop {}
  {\bibfield  {journal} {\bibinfo  {journal} {Reviews of modern physics}\
  }\textbf {\bibinfo {volume} {82}},\ \bibinfo {pages} {1539} (\bibinfo {year}
  {2010})}\BibitemShut {NoStop}%
\bibitem [{\citenamefont {Ruderman}\ and\ \citenamefont
  {Kittel}(1954)}]{ruderman1954indirect}%
  \BibitemOpen
  \bibfield  {author} {\bibinfo {author} {\bibfnamefont {M.~A.}\ \bibnamefont
  {Ruderman}}\ and\ \bibinfo {author} {\bibfnamefont {C.}~\bibnamefont
  {Kittel}},\ }\href@noop {} {\bibfield  {journal} {\bibinfo  {journal}
  {Physical Review}\ }\textbf {\bibinfo {volume} {96}},\ \bibinfo {pages} {99}
  (\bibinfo {year} {1954})}\BibitemShut {NoStop}%
\bibitem [{\citenamefont {Kasuya}(1956)}]{kasuya1956theory}%
  \BibitemOpen
  \bibfield  {author} {\bibinfo {author} {\bibfnamefont {T.}~\bibnamefont
  {Kasuya}},\ }\href@noop {} {\bibfield  {journal} {\bibinfo  {journal}
  {Progress of theoretical physics}\ }\textbf {\bibinfo {volume} {16}},\
  \bibinfo {pages} {45} (\bibinfo {year} {1956})}\BibitemShut {NoStop}%
\bibitem [{\citenamefont {Yosida}(1957)}]{yosida1957magnetic}%
  \BibitemOpen
  \bibfield  {author} {\bibinfo {author} {\bibfnamefont {K.}~\bibnamefont
  {Yosida}},\ }\href@noop {} {\bibfield  {journal} {\bibinfo  {journal}
  {Physical Review}\ }\textbf {\bibinfo {volume} {106}},\ \bibinfo {pages}
  {893} (\bibinfo {year} {1957})}\BibitemShut {NoStop}%
\bibitem [{\citenamefont {Hayami}\ and\ \citenamefont
  {Motome}(2021)}]{hayami2021charge}%
  \BibitemOpen
  \bibfield  {author} {\bibinfo {author} {\bibfnamefont {S.}~\bibnamefont
  {Hayami}}\ and\ \bibinfo {author} {\bibfnamefont {Y.}~\bibnamefont
  {Motome}},\ }\href@noop {} {\bibfield  {journal} {\bibinfo  {journal} {arXiv
  preprint arXiv:2108.04997}\ } (\bibinfo {year} {2021})}\BibitemShut {NoStop}%
\bibitem [{\citenamefont {Shimomura}\ \emph {et~al.}(2019)\citenamefont
  {Shimomura}, \citenamefont {Murao}, \citenamefont {Tsutsui}, \citenamefont
  {Nakao}, \citenamefont {Nakamura}, \citenamefont {Hedo}, \citenamefont
  {Nakama},\ and\ \citenamefont {{\=O}nuki}}]{shimomura2019lattice}%
  \BibitemOpen
  \bibfield  {author} {\bibinfo {author} {\bibfnamefont {S.}~\bibnamefont
  {Shimomura}}, \bibinfo {author} {\bibfnamefont {H.}~\bibnamefont {Murao}},
  \bibinfo {author} {\bibfnamefont {S.}~\bibnamefont {Tsutsui}}, \bibinfo
  {author} {\bibfnamefont {H.}~\bibnamefont {Nakao}}, \bibinfo {author}
  {\bibfnamefont {A.}~\bibnamefont {Nakamura}}, \bibinfo {author}
  {\bibfnamefont {M.}~\bibnamefont {Hedo}}, \bibinfo {author} {\bibfnamefont
  {T.}~\bibnamefont {Nakama}}, \ and\ \bibinfo {author} {\bibfnamefont
  {Y.}~\bibnamefont {{\=O}nuki}},\ }\href@noop {} {\bibfield  {journal}
  {\bibinfo  {journal} {Journal of the Physical Society of Japan}\ }\textbf
  {\bibinfo {volume} {88}},\ \bibinfo {pages} {014602} (\bibinfo {year}
  {2019})}\BibitemShut {NoStop}%
\bibitem [{\citenamefont {Kaneko}\ \emph {et~al.}(2021)\citenamefont {Kaneko},
  \citenamefont {Kawasaki}, \citenamefont {Nakamura}, \citenamefont {Munakata},
  \citenamefont {Nakao}, \citenamefont {Hanashima}, \citenamefont {Kiyanagi},
  \citenamefont {Ohhara}, \citenamefont {Hedo}, \citenamefont {Nakama} \emph
  {et~al.}}]{kaneko2021charge}%
  \BibitemOpen
  \bibfield  {author} {\bibinfo {author} {\bibfnamefont {K.}~\bibnamefont
  {Kaneko}}, \bibinfo {author} {\bibfnamefont {T.}~\bibnamefont {Kawasaki}},
  \bibinfo {author} {\bibfnamefont {A.}~\bibnamefont {Nakamura}}, \bibinfo
  {author} {\bibfnamefont {K.}~\bibnamefont {Munakata}}, \bibinfo {author}
  {\bibfnamefont {A.}~\bibnamefont {Nakao}}, \bibinfo {author} {\bibfnamefont
  {T.}~\bibnamefont {Hanashima}}, \bibinfo {author} {\bibfnamefont
  {R.}~\bibnamefont {Kiyanagi}}, \bibinfo {author} {\bibfnamefont
  {T.}~\bibnamefont {Ohhara}}, \bibinfo {author} {\bibfnamefont
  {M.}~\bibnamefont {Hedo}}, \bibinfo {author} {\bibfnamefont {T.}~\bibnamefont
  {Nakama}},  \emph {et~al.},\ }\href@noop {} {\bibfield  {journal} {\bibinfo
  {journal} {Journal of the Physical Society of Japan}\ }\textbf {\bibinfo
  {volume} {90}},\ \bibinfo {pages} {064704} (\bibinfo {year}
  {2021})}\BibitemShut {NoStop}%
\bibitem [{\citenamefont {Shang}\ \emph {et~al.}(2021)\citenamefont {Shang},
  \citenamefont {Xu}, \citenamefont {Gawryluk}, \citenamefont {Ma},
  \citenamefont {Shiroka}, \citenamefont {Shi},\ and\ \citenamefont
  {Pomjakushina}}]{shang2021anomalous}%
  \BibitemOpen
  \bibfield  {author} {\bibinfo {author} {\bibfnamefont {T.}~\bibnamefont
  {Shang}}, \bibinfo {author} {\bibfnamefont {Y.}~\bibnamefont {Xu}}, \bibinfo
  {author} {\bibfnamefont {D.}~\bibnamefont {Gawryluk}}, \bibinfo {author}
  {\bibfnamefont {J.}~\bibnamefont {Ma}}, \bibinfo {author} {\bibfnamefont
  {T.}~\bibnamefont {Shiroka}}, \bibinfo {author} {\bibfnamefont
  {M.}~\bibnamefont {Shi}}, \ and\ \bibinfo {author} {\bibfnamefont
  {E.}~\bibnamefont {Pomjakushina}},\ }\href@noop {} {\bibfield  {journal}
  {\bibinfo  {journal} {Physical Review B}\ }\textbf {\bibinfo {volume}
  {103}},\ \bibinfo {pages} {L020405} (\bibinfo {year} {2021})}\BibitemShut
  {NoStop}%
\bibitem [{\citenamefont {Kawasaki}\ \emph {et~al.}(2016)\citenamefont
  {Kawasaki}, \citenamefont {Kaneko}, \citenamefont {Nakamura}, \citenamefont
  {Aso}, \citenamefont {Hedo}, \citenamefont {Nakama}, \citenamefont {Ohhara},
  \citenamefont {Kiyanagi}, \citenamefont {Oikawa}, \citenamefont {Tamura}
  \emph {et~al.}}]{kawasaki2016magnetic}%
  \BibitemOpen
  \bibfield  {author} {\bibinfo {author} {\bibfnamefont {T.}~\bibnamefont
  {Kawasaki}}, \bibinfo {author} {\bibfnamefont {K.}~\bibnamefont {Kaneko}},
  \bibinfo {author} {\bibfnamefont {A.}~\bibnamefont {Nakamura}}, \bibinfo
  {author} {\bibfnamefont {N.}~\bibnamefont {Aso}}, \bibinfo {author}
  {\bibfnamefont {M.}~\bibnamefont {Hedo}}, \bibinfo {author} {\bibfnamefont
  {T.}~\bibnamefont {Nakama}}, \bibinfo {author} {\bibfnamefont
  {T.}~\bibnamefont {Ohhara}}, \bibinfo {author} {\bibfnamefont
  {R.}~\bibnamefont {Kiyanagi}}, \bibinfo {author} {\bibfnamefont
  {K.}~\bibnamefont {Oikawa}}, \bibinfo {author} {\bibfnamefont
  {I.}~\bibnamefont {Tamura}},  \emph {et~al.},\ }\href@noop {} {\bibfield
  {journal} {\bibinfo  {journal} {Journal of the Physical Society of Japan}\
  }\textbf {\bibinfo {volume} {85}},\ \bibinfo {pages} {114711} (\bibinfo
  {year} {2016})}\BibitemShut {NoStop}%
\bibitem [{\citenamefont {Nakamura}\ \emph {et~al.}(2015)\citenamefont
  {Nakamura}, \citenamefont {Uejo}, \citenamefont {Honda}, \citenamefont
  {Takeuchi}, \citenamefont {Harima}, \citenamefont {Yamamoto}, \citenamefont
  {Haga}, \citenamefont {Matsubayashi}, \citenamefont {Uwatoko}, \citenamefont
  {Hedo} \emph {et~al.}}]{nakamura2015transport}%
  \BibitemOpen
  \bibfield  {author} {\bibinfo {author} {\bibfnamefont {A.}~\bibnamefont
  {Nakamura}}, \bibinfo {author} {\bibfnamefont {T.}~\bibnamefont {Uejo}},
  \bibinfo {author} {\bibfnamefont {F.}~\bibnamefont {Honda}}, \bibinfo
  {author} {\bibfnamefont {T.}~\bibnamefont {Takeuchi}}, \bibinfo {author}
  {\bibfnamefont {H.}~\bibnamefont {Harima}}, \bibinfo {author} {\bibfnamefont
  {E.}~\bibnamefont {Yamamoto}}, \bibinfo {author} {\bibfnamefont
  {Y.}~\bibnamefont {Haga}}, \bibinfo {author} {\bibfnamefont {K.}~\bibnamefont
  {Matsubayashi}}, \bibinfo {author} {\bibfnamefont {Y.}~\bibnamefont
  {Uwatoko}}, \bibinfo {author} {\bibfnamefont {M.}~\bibnamefont {Hedo}},
  \emph {et~al.},\ }\href@noop {} {\bibfield  {journal} {\bibinfo  {journal}
  {Journal of the Physical Society of Japan}\ }\textbf {\bibinfo {volume}
  {84}},\ \bibinfo {pages} {124711} (\bibinfo {year} {2015})}\BibitemShut
  {NoStop}%
\bibitem [{\citenamefont {Allen}\ and\ \citenamefont
  {et~al}(2021)}]{Kevin2021}%
  \BibitemOpen
  \bibfield  {author} {\bibinfo {author} {\bibfnamefont {K.}~\bibnamefont
  {Allen}}\ and\ \bibinfo {author} {\bibnamefont {et~al}},\ }\href@noop {}
  {\bibfield  {journal} {\bibinfo  {journal} {In Preperation}\ } (\bibinfo
  {year} {2021})}\BibitemShut {NoStop}%
\end{thebibliography}%

\cleardoublepage

%%%%%%%%%% Merge with supplemental materials %%%%%%%%%%
\pagebreak
\widetext
\begin{center}
\textbf{\large Supplemental Materials:\\ Incommensurate magnetic orders and topological Hall effect in the square-net centrosymmetric EuGa$_2$Al$_2$ system}

\end{center}
%%%%%%%%%% Merge with supplemental materials %%%%%%%%%%
%%%%%%%%%% Prefix a "S" to all equations, figures, tables and reset the counter %%%%%%%%%%
\setcounter{equation}{0}
\setcounter{figure}{0}
\setcounter{table}{0}
\setcounter{page}{1}
\makeatletter
\renewcommand{\theequation}{S\arabic{equation}}
\renewcommand{\thefigure}{S\arabic{figure}}

\section{Neutron data}

 \begin{figure}[h]
  \includegraphics[width=16cm,height=10cm]{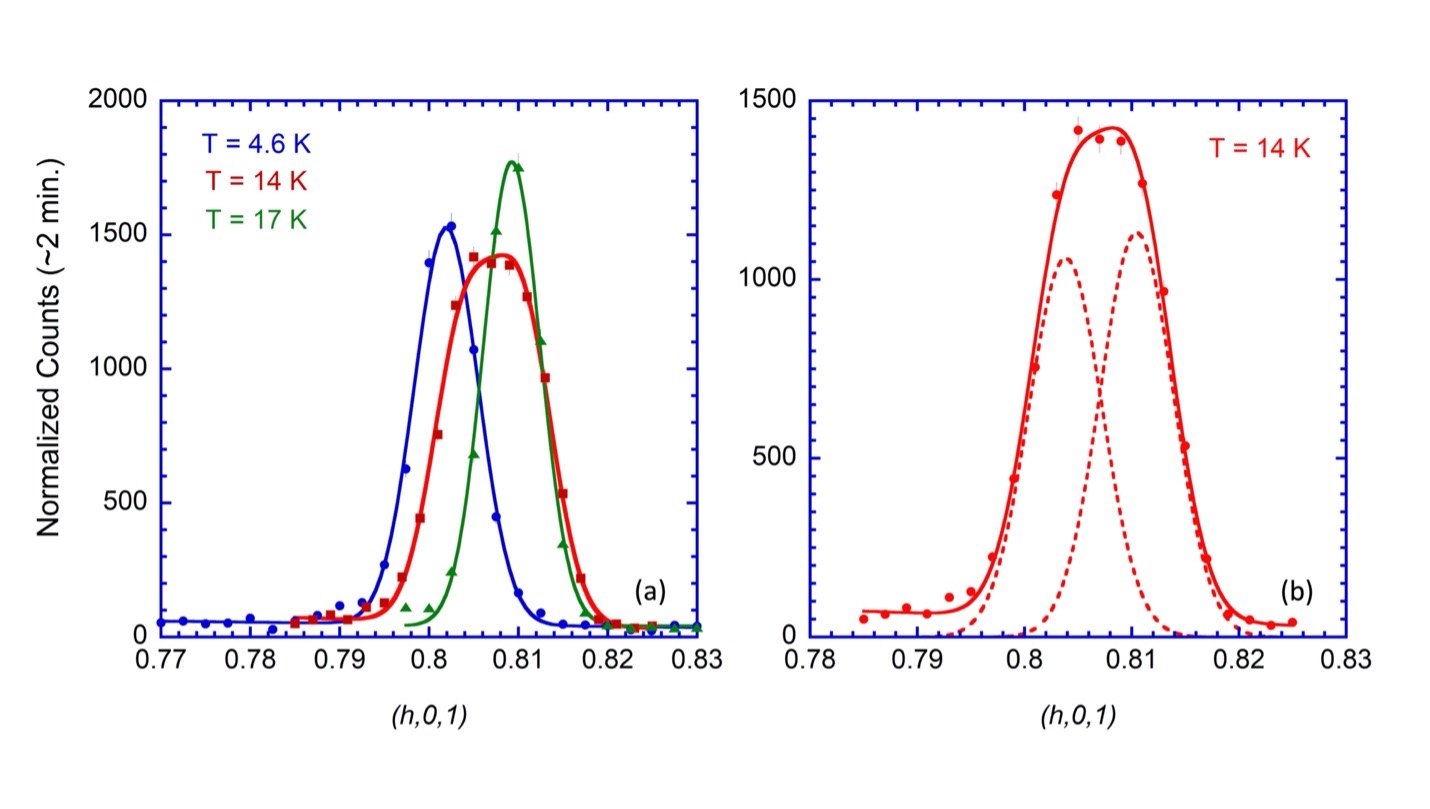}
\caption{(a) High resolution neutron diffraction scans taken on HB-3 along ($h,0,1$) in each of the three phase regimes AFM3, AFM2, and AFM1 at $T$ = 4.6, 14, 17 K, respectively. (b) The scan in the AFM2 regime has the largest integrated intensity because of its breadth, and can be fit with two resolution-limited Gaussian peaks, indicating that the two types of magnetic order coexist in this mixed regime with a first-order transition between the high temperature helical and low temperature cycloid phases.}
\label{nuet2}
\end{figure}
In the AFM2 phase the width of the peak exhibits some broadening even when employing coarse resolution.  We therefore tightened the collimation to the best available at the time, 48'-20'-20'-70', and those data are shown in Fig. \ref{nuet2}.  In the AFM1 and AFM3 phases the peaks are resolution limited, while the broader peak in the AFM2 regime can be fit with two resolution-limited peaks, indicating a first-order transition between AFM1 and AFM3 phases that overlap.  We also carried out an experiment using the Si(111) monochromator with 9 meV incident energy and tight collimation, but given the limited statistics in that configuration it was not possible to obtain improved data.

 \begin{figure}
  \includegraphics[width=10cm]{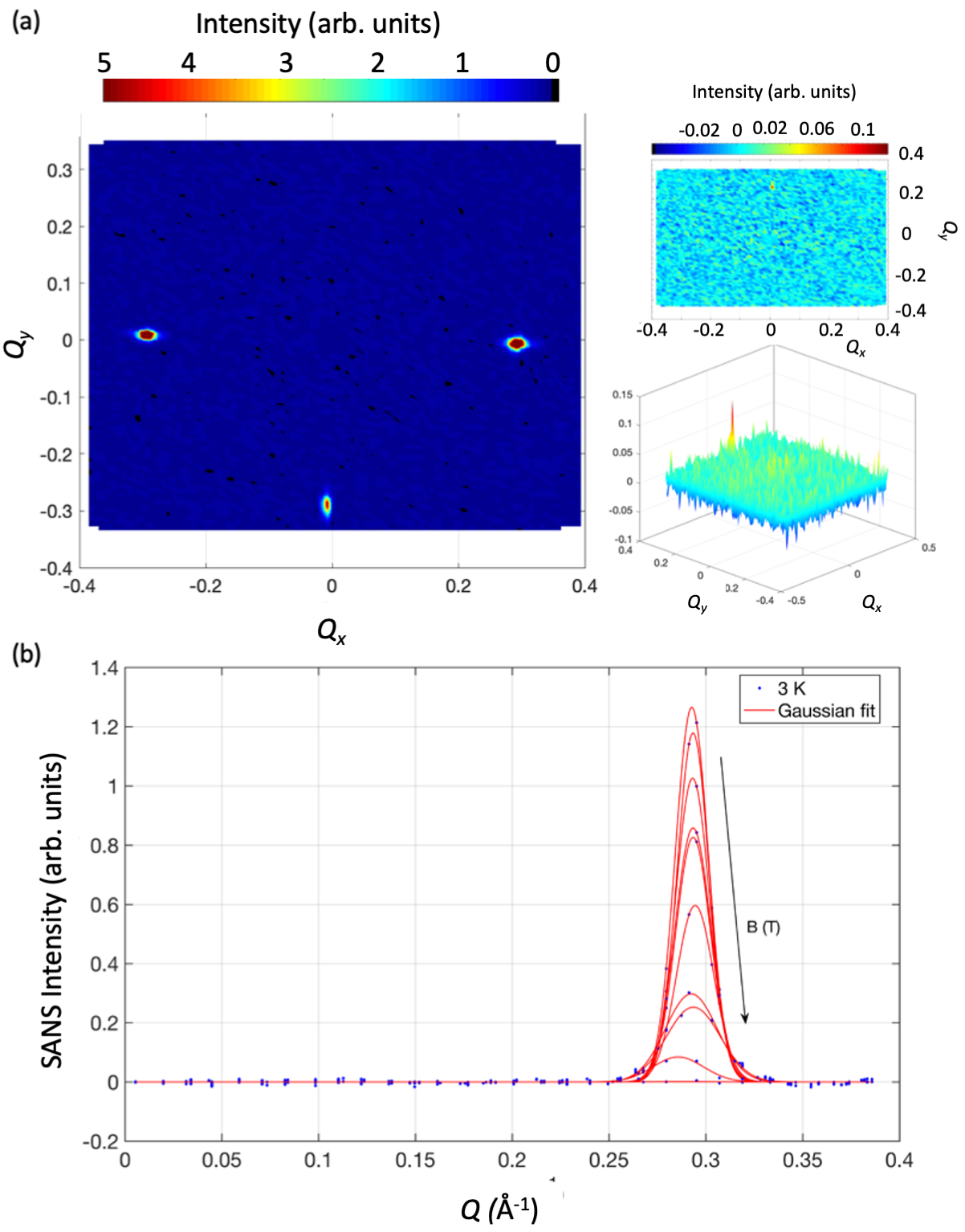}
\caption{(a) Small-angle neutron diffraction (SANS) pattern of independently measured incommensurate magnetic peaks in the ($h$,$k$,0) plane shows tetragonally equivalent magnetic wavevectors. Each of the tetragonally equivalent peaks were measured independently, subtracted from the background scan at 25 K, and summed to give the resulting pattern. Inset: Due to the limited tilt angle of the magnet, the measurement only catches the Bragg tail of the peak along +$Q_y$, which results in a much reduced intensity. (b) Line profile of the SANS intensity for the incommensurate wavevector along -$Q_x$ for magnetic fields ranging from $\mu_0H$ = 0 – 3 T. Solid lines are Gaussian fits to the data. Extracted parameters as a function of field are shown in Fig.~\ref{neutron3}  of the main text.}
\label{fieldneutSM}
\end{figure}

\cleardoublepage
\pagebreak
\section{Temperature dependent X-ray diffraction data}

 \begin{figure}[h]
  \includegraphics[width=\linewidth]{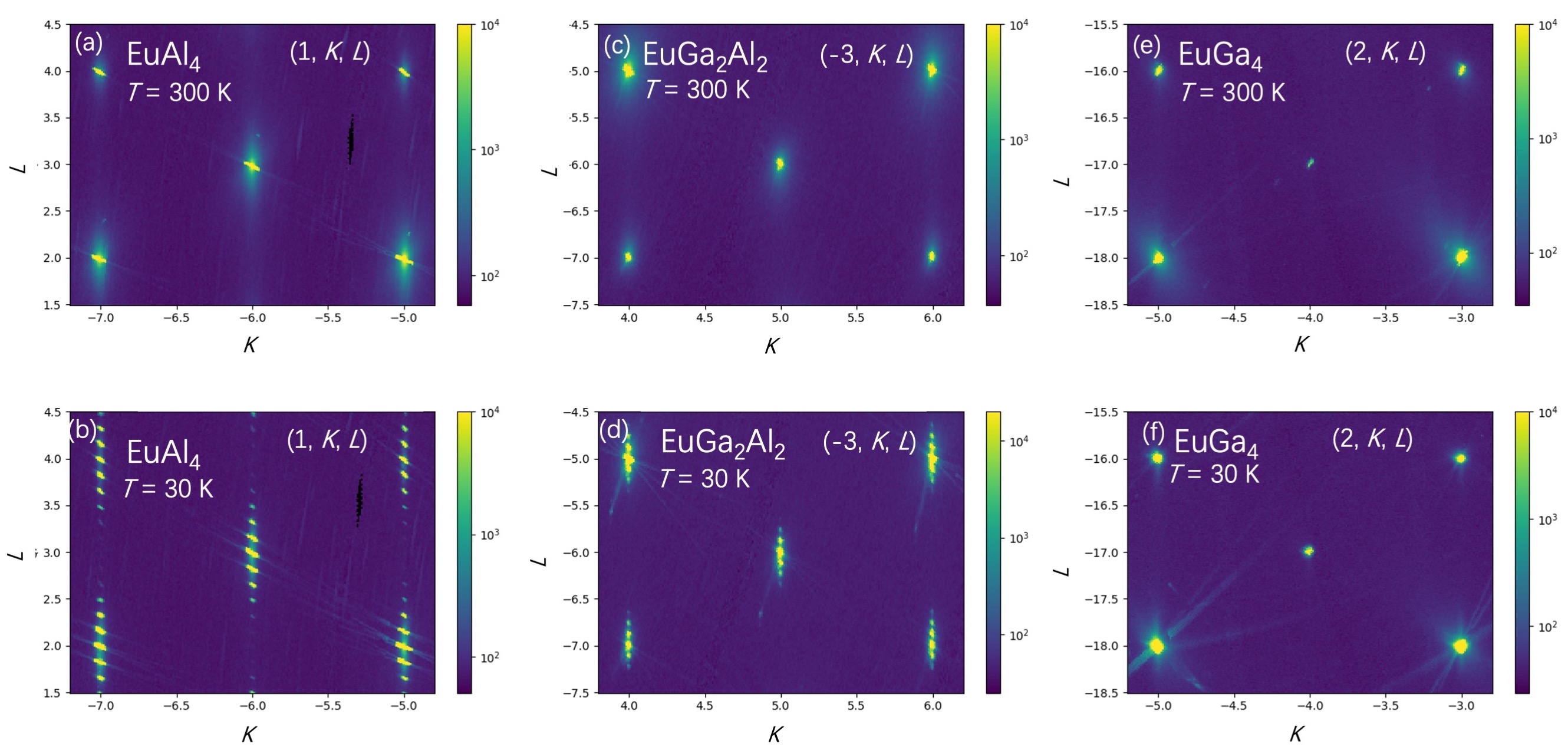}
\caption{X-ray diffraction reciprocal space maps. Reciprocal space maps in the (1, $K$, $L$) scattering plane of EuAl$_4$ at (a) 300 K and (b) 30 K. Clear superlattice reflections are observed forming along the (0,0,$L$) direction at 30 K. Reciprocal space maps in the (-3, $K$, $L$) scattering plane of EuGa$_2$Al$_2$ at (c) 300 K and (d) 30 K. Again superlattice reflections are observed forming along the (0,0,$L$) direction at 30 K. Reciprocal space maps in the (2, $K$, $L$) scattering plane of EuGa$_4$ at (e) 300 K and (f) 30 K. No superlattice reflections are observed down to 30 K. }
\label{XRD1}
\end{figure}

 \begin{figure}
  \includegraphics[width=\linewidth]{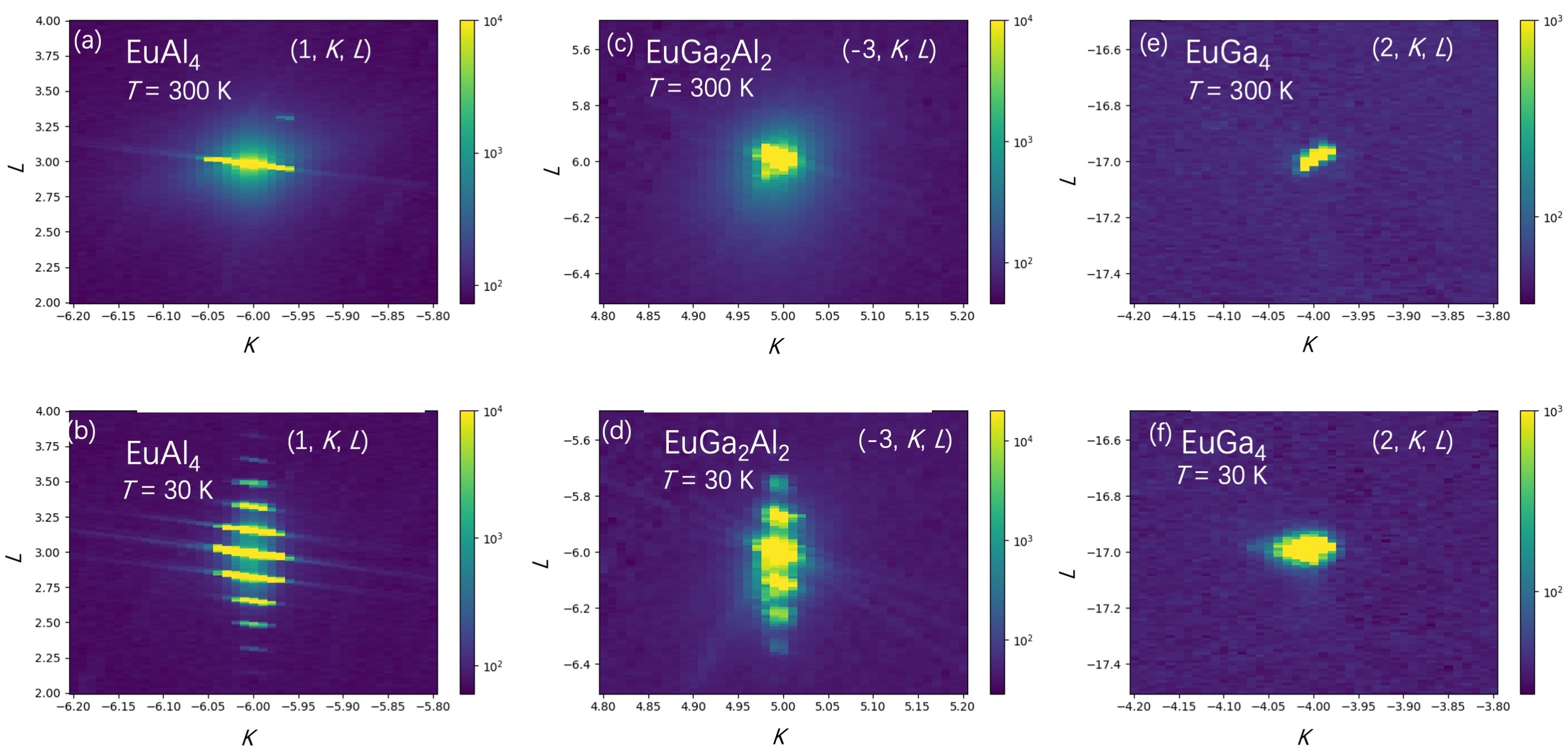}
\caption{High resolution X-ray diffraction reciprocal space maps corresponding to Fig. \ref{XRD1} for EuAl$_4$ at 300~K and 30~K (a,b)  EuGa$_2$Al$_2$ at 300~K and 30~K (c,d), and EuGa$_4$ at 300~K and 30~K (e,f).}
\label{XRD2}
\end{figure}

 \begin{figure}
  \includegraphics[width=\linewidth]{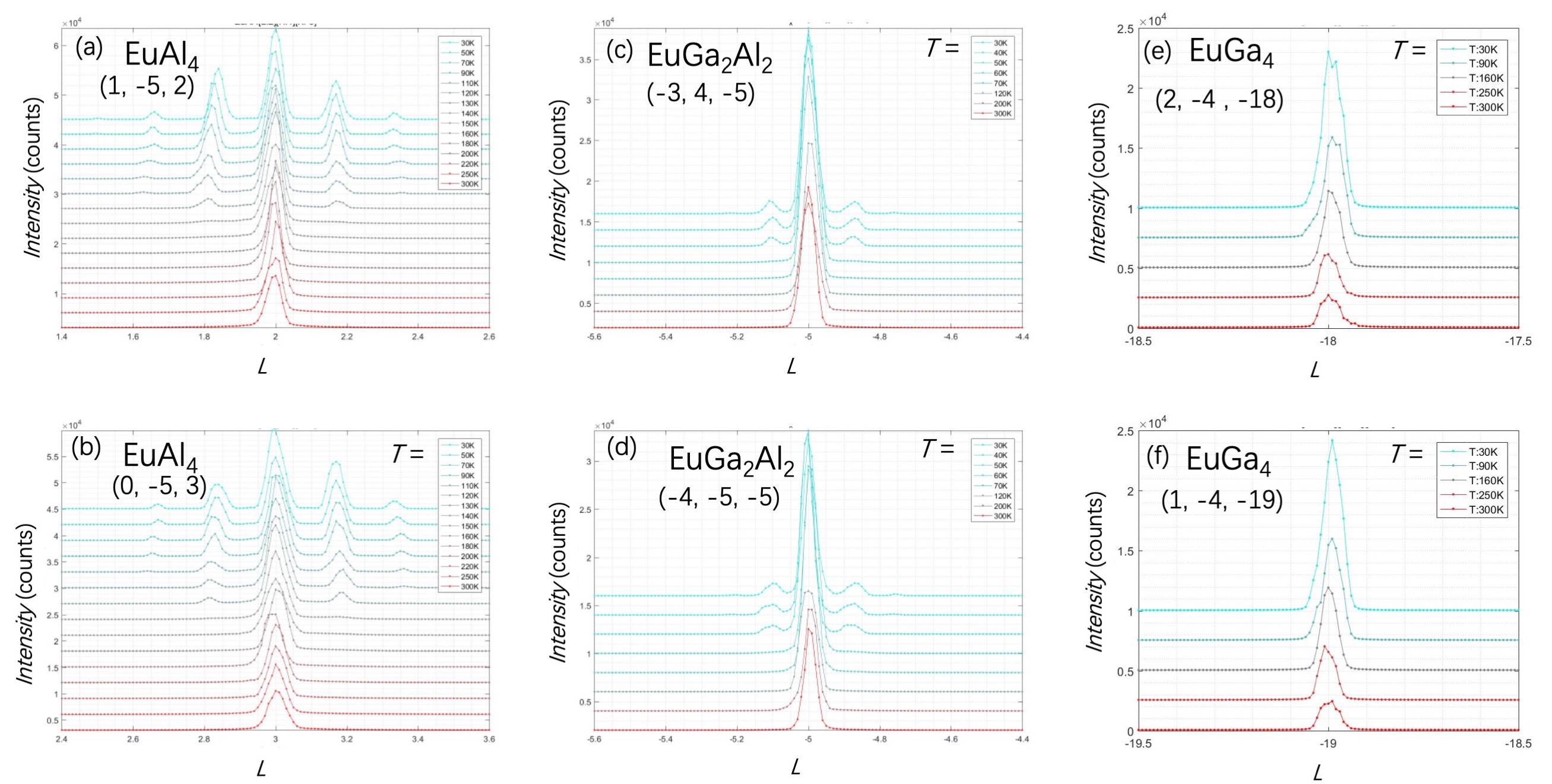}
\caption{Temperature-dependent line cuts of high resolution X-ray diffraction data between 300 and 30 K around the (a) (1, -5, 2) nuclear Bragg peak and (b) (0,-5,3) nuclear Bragg peak of EuAl$_4$. Superlattice reflections emerge near the $T$ = 140 K providing evidence for CDW behavior consistent with previous reports \cite{shimomura2019lattice, kaneko2021charge}. Line cuts near the (c) (-3,4,-5) and (d) (-4,-5,-5) nuclear Bragg peaks of EuGa$_2$Al$_2$ giving evidence for CDW behavior discussed in the main text. Line cuts near the (e) (2, -4, -18) and  (f) (1,-4,-19) nuclear Bragg peak of EuGa$_4$. No evidence of superlattice reflections are observed and hence no evidence for a CDW is observed for temperatures above 30 K. }
\label{XRD3}
\end{figure}

\end{document}